\documentclass[12pt,preprint]{aastex}
\usepackage{natbib}
\usepackage{bm}
\usepackage{adjustbox,lipsum}
\usepackage{enumerate}
\usepackage{hyperref}
\usepackage{float}
\usepackage[latin1]{inputenc}
\newcommand{\brac}[1]{\langle #1 \rangle}
\newcommand{\pd}{\partial}

\newcommand{\mean}[1]{\overline{#1}}

\def\Ra{\mbox{\rm Ra}}
\def\Ma{\mbox{\rm Ma}}

\def\Ro{\mbox{\rm Ro}}

\def\cp{c_{\rm p}}

\def\Rs{R_{\odot}}
\def\cs{c_{\rm s}}

\def\urms{u_{\rm rms}}

\def\etat{\eta_{\rm t}}
\def\etato{\eta_{\rm t0}}

\usepackage{graphicx}
\usepackage{graphics}
\graphicspath{{fig/}{png/}}

\slugcomment{Not to appear in Nonlearned J., 45.}

\shorttitle{Tachocline in TP balance}
\shortauthors{Guerrero et al.}
\begin{document}

\title{On the role of tachoclines in solar and stellar dynamos}

\author{G. Guerrero}
\affil{Physics Department, Universidade Federal de Minas Gerais,
Av. Antonio Carlos, 6627, Belo Horizonte, MG, Brazil, 31270-901}
\email{guerrero@fisica.ufmg.br}

\author{P.~K. Smolarkiewicz}
\affil{European Centre for Medium-Range Weather Forecasts, Reading RG2 9AX, UK}
\email{smolar@ecmwf.int}

\author{E. M. de Gouveia Dal Pino}
\affil{Astronomy Department, IAG-USP 
Rua do matão, 1226, São Paulo, SP, Brasil, 05508-090}
\email{dalpino@astro.iag.usp.br}

\author{A.~G. Kosovichev}
\affil{New Jersey Institute of Technology, Newark, NJ 07103,USA}
\email{sasha@bbso.njit.edu}

\and
\author{N. N. Mansour}
\affil{NASA, Ames Research Center, Moffett Field, Mountain View, CA 94040, USA}
\email{Nagi.N.Mansour@nasa.gov}

\begin{abstract}
Rotational shear layers at the boundary between radiative and convective
zones, tachoclines,  play a key role in the process of magnetic field generation
in solar-like stars. We present two sets of global simulations 
of rotating turbulent convection and dynamo. The first set considers a stellar
convective envelope only; the second one, aiming at the formation of a tachocline,
considers  also the upper part of the radiative zone. Our results indicate that
the resulting mean-flows and dynamo properties like the growth rate, 
saturation energy and mode depend on the Rossby ($\Ro$) number. 
For the first set of models either oscillatory (with $\sim 2$ yr period) 
or steady dynamo solutions are obtained. The models in the second set naturally 
develop a tachocline which, in turn, 
leads to the generation of strong mean magnetic field.
Since the field is also deposited into the stable deeper layer,
its evolutionary time-scale is much longer than in the models without 
a tachocline. Surprisingly, the magnetic field in the upper turbulent 
convection zone evolves in the same time scale as the deep field. These models 
result in either an oscillatory dynamo with $\sim 30$ yr period or in a
steady dynamo depending on $\Ro$. In terms of the mean-field
dynamo coefficients computed using FOSA, the field evolution in the 
oscillatory models without a tachocline seems to be consistent 
with dynamo waves propagating according to the Parker-Yoshimura sign 
rule. In the models with tachoclines the dynamics is more complex  
involving other transport mechanisms as well as tachocline instabilities.
\end{abstract}


\section{Introduction}
One of the most challenging questions in astrophysics is
the origin of the stellar magnetism. More specifically, we do not 
know yet where inside the stars the large-scale magnetic 
fields are generated and sustained. The best astrophysics 
laboratory to study
stellar magnetism is our host star, the Sun. Its magnetic field
at photospheric levels can be observed in detail, and is 
depicted in the well known time-latitude ``butterfly" diagram. It summarizes 
the cyclic properties of the surface evolution of the field, 
its migration patterns, the periodicity of magnetic activity and 
polarity reversals. The morphology and distribution of the field in
deeper layers is, unfortunately, still evasive to any kind of
observation. In other solar-like stars (in the main sequence and
with spectral types from F to M)  the observations
of magnetic field are less detailed. Nevertheless, cyclic
magnetic activity, with cycle-periods between 5 and 25 years,
has been observed in a good sample of stars \citep{Baliunas+95}.
In the same work, the authors reported  also observations of 
flat, non-oscillatory, 
magnetic activity. A general trend
is that the strength of the magnetic field increases with the
rotation rate of the star. More recently, \cite{Petit+08} have been able
to infer the topology of the surface magnetic field of four stars
which could be classified as solar twins (with stellar parameters
close to those of the Sun). Their results indicate that the 
toroidal field strength is proportional to the rotation rate
but the poloidal component anti-correlates with it. All these
observations allow us to look at the stellar magnetism from a 
broad perspective which can add new constraints in the underlying 
processes and ultimately lead to a better understanding of it.

From the theoretical point of view, the generation and evolution
of large-scale magnetic fields in stars (and other cosmic objects) has
been studied using the mean-field dynamo theory \citep[see][for a complete 
review]{BS05}.  The mean-field induction equation implies that the 
dominant term generating an azimuthal (toroidal) magnetic field 
depends on gradients of the angular velocity, 
$ (\mean{\bm B}_p \cdot \nabla) \Omega$, 
where $\Omega$ is the mean angular velocity, and
$\mean{\bm B}_p$ is a pre-existing large-scale poloidal field. The 
generation of  the poloidal field relies on the turbulent helicities, 
the so-called $\alpha$-effect,  a non-diffusive contribution of the
electromotive force whose dependence on the convective turbulence 
and rotation is still uncertain.  Under appropriate conditions, 
the evolution of the magnetic field in this theory is consistent with 
dynamo waves that, according to the \cite{Pa55}-\cite{Y75} sign rule,
propagate in the direction of:
\begin{equation}
\label{eq.py}
{\bm s}=\alpha\nabla\Omega \times \hat{\bm e}_{\phi},
\end{equation}
where $\hat{\bm e}_{\phi}$ is the unit vector in the azimuthal direction.
The validity of this rule has been extensively tested in 
two-dimensional kinematic mean-field models \cite[see, e.g.][]{Ch10}. 
For the solar rotation inferred from helioseismology \citep{SCHOU98}, 
the signs of the 
$\nabla \Omega$ components are well known. The sign of $\alpha$ estimated 
for granular convection is positive (negative) in the 
Northern (Southern) hemisphere.   These properties favour a dynamo process
distributed over the entire convection zone with the photospheric
evolution shaped by the near-surface shear layer (NSSL) \citep{B05,PK11}. 
Alternative, flux-transport dynamo models in which
the surface evolution is shaped by the dynamics at the
tachocline (and magnetic buoyancy) require coherent meridional 
motions directed equatorward at, or slightly below the tachocline.

Different observational techniques have revealed a well-organized, 
though fluctuating, poleward meridional flow at the solar surface 
\citep[e.g.,][]{UL10,Ha12} with amplitudes of about $15$~ms$^{-1}$. 
However, recent helioseismology inferences have determined the existence 
of two or more cells in the radial direction \citep{zbkdh13,str13}. Thus, the existence 
of a coherent equatorial flow, able to transport the magnetic flux 
from the poles to the latitudes of sunspot activity, is unlikely. 

The dynamo wave propagation sign rule remains valid also for mean-field 
models considering the dynamical evolution of the $\alpha$-effect \citep{GCB10}.
In these models the kinetic component of the $\alpha$-effect changes  
in time due to the backreaction of the magnetic field on the plasma motion. 
The magnetic $\alpha$-effect depends on
the distribution of the small-scale current helicity, and is a consequence
of the magnetic helicity conservation \citep{PFL76}.  It is regularly
assumed that it diminishes the inductive action of the kinetic helicity 
\citep[e.g.,][]{SSS07,GCB10,MMTB11}. For high magnetic Reynold numbers, this 
contribution can lead to the total suppression of the dynamo
action, the so-called catastrophic quenching, in the case when there is no
effective mechanism to remove the small-scale current helicity \citep{GCB10}.
Recent works \citep{VS14}, and the results that will be presented 
here, indicate that this is not necessarily the only case, and that
the magnetic contribution to the $\alpha$-effect could be a source 
of magnetic field. A similar idea is suggested by \cite{Bonanno13}.

Several kinematic or dynamic mean-field models have qualitatively 
reproduced the observed surface features of the solar magnetic field
\citep{DG01,CNC04,GDP08,PK11,PK13}. Based on these models, \cite{JBB10,P14} have
also studied dynamos for solar-like stars with the aim of 
reproducing observed trends for magnetic field strenghts and cycle periods. 
Unfortunately, parameters like
helicity and/or turbulent diffusion, in general, are poorly determined,
and thus the physics of the solar and stellar dynamos cannot be described 
unambiguously by the mean-field models. An alternative approach is 
provided by global 3D MHD simulations. In this class of modeling, even 
though the parametric regime is still far from the conditions of the stellar
interiors, the physics is self-consistently described and provides
an important insight into the turbulent dynamics and dynamo.

For instance, systematic hydrodynamic (HD) studies of rotating turbulent
convection, for the Sun and solar-like stars, have found that the 
Rossby number, $\Ro=\urms/2\Omega_0 L$ (where $\Omega_0$ is the frame
rotation rate, $L$ the characteristic large-scale length of the flow 
and $\urms$ is the typical turbulent velocity), characterizes the resulting 
mean flows \citep{Gi76,GG82,SF07,MDBB11,KMGBC11,GSKM13b,GYMRW14,FM15}. 
For large values of $\Ro$ the models result in an
anti-solar differential rotation (with faster rotating
poles and a slower rotating equatorial region), and a meridional
circulation with single cells per hemisphere. In the models with small 
values of $\Ro$
the differential rotation is solar-like, and the meridional flow consists of
 several circulation cells per hemisphere.  The transition 
between these regimes is sharp, occurring in a narrow range of $\Ro$.  Several
properties of the observed solar differential rotation, like the latitudinal
dependence of the angular velocity, the tachocline \citep{GCS10}, and a NSSL, 
although so far only at high latitudes \citep{GSKM13b,HR15}, have been 
successfully reproduced. 

Magnetohydrodynamic (MHD) global dynamo  models, which have been
developed for more than a decade \citep[e.g.,][]{BMT04}, have provided
a wide spectrum of results, but the understanding of the dependence of the 
mean-flows or magnetic fields on stellar parameters
is still incomplete. Furthermore, to our knowledge, no MHD model so 
far has been able to reproduce the solar differential rotation.

MHD simulations of fast-rotating stars by \cite{BBBMT10} have been 
able to obtain large
scale dynamo action. Their results revealed a steady magnetic field
organized in the shape of torus, or wreaths, around the 
equator and with opposite polarity accros the hemispheres. Non-oscillatory
dynamo solutions have also been obtained by \cite{SKB15}
Oscillatory dynamo solutions in the global MHD models were first obtained 
by \cite{GCS10} using the EULAG-MHD code with an implicit sub-grid scale (SGS)  
formulation.  Since then other groups have been able to simulate 
magnetic cycles with the use of SGS turbulent models 
\citep{ABT14}, or also in higher-resolution simulations 
by considering
only a fraction (a wedge) of the star \citep{KMB12,WKKB14}.  Although the
dynamo regimes of these models are not the same, 
it is noteworthy that the cycle periods obtained in the model of 
\cite{GCS10} and the models of \citep{KMB12, ABT14, WKKB14} are 
rather different. As it will be shown later
in this paper, this difference may be explained by the absence of a
tachocline and a radiative stable layer in the models of 
\cite{KMB12, ABT14, WKKB14}.  By considering
a forcing function to impose a constant angular velocity at the 
bottom of the domain, \citep{BMBT06} analyzed  the effects of 
a tachocline in 
convective dynamo simulations. Although they found strong toroidal 
magnetic field development at this imposed shear layer, 
they  observed  no field reversals. \cite{MYK13} explored the effects
of penetrative convection in spherical dynamo simulations. 
Due to dissipative effects the tachocline obtained in their model
is not well defined. However, they obtained stronger and cyclic 
large-scale toroidal magnetic field, demonstrating the importance of the
stable layer in the storage of the magnetic field.

In this paper we compare 3D global MHD convective dynamo 
simulations with different Rossby numbers for models with and 
without the tachocline. Our goal is to 
compare the global properties of the mean-flows and the resultant
magnetic activity, and contrast them with observational signatures.
Of particular interest are the inductive and diffusive terms 
resulting from the differential rotation and the collective 
effects of turbulence.  
Altogether, this will allow us to elucidate on the
importance of the tachocline in solar and stellar dynamos.

\section{The model}
We adopt a full spherical shell, $0\le \phi \le 2\pi$, 
$0\le \theta \le \pi$,  the radial domain has its bottom boundary at $r_b=0.61\Rs$ 
for the models that develop a tachocline, and $r_b=0.72\Rs$ for the models
without it, and the upper boundary at $r_t=0.96\Rs$.
Unlike \cite{GSKM13b}, where the anelastic equations of \cite{LH82} were
employed, here we solve their MHD extension \citep{GCS10}:
\begin{equation}                                                                                               
{\bm \nabla}\cdot(\rho_s\bm u)=0, \label{equ:cont}
\end{equation}
\begin{equation}
	\frac{D \bm u}{Dt}+ 2{\bm \Omega} \times {\bm u} =  
    -{\bm \nabla}\left(\frac{p'}{\rho_s}\right) + {\bm g}\frac{\Theta'}
     {\Theta_s} + \frac{1}{\mu_0 \rho_s}({\bm B} \cdot \nabla) {\bm B} \;, \label{equ:mom} 
\end{equation}
\begin{equation}
	\frac{D \Theta'}{Dt} = -{\bm u}\cdot {\bm \nabla}\Theta_e -\frac{\Theta'}{\tau}\;, \label{equ:en} 
\end{equation}
\begin{equation}
 \frac{D {\bm B}}{Dt} = ({\bm B}\cdot \nabla) {\bm u} - {\bm B}(\nabla \cdot {\bm u})  \;,
 \label{equ:in} 
\end{equation}
\noindent
where $D/Dt = \pd/\pd t + \bm{u} \cdot {\bm \nabla}$ is the total
time derivative, ${\bm u}$ is the velocity field in a rotating 
frame with ${\bm \Omega}=\Omega_0(\cos\theta,-\sin\theta,0)$,
$p'$ is a pressure perturbation variable that accounts for both the gas 
and magnetic pressure, 
${\bm B}$ is the magnetic field, and $\Theta'$ is the potential temperature 
perturbation with respect to an
ambient state $\Theta_e$ \citep[see \S3 of][for a comprehensive discussion]{GSKM13b}.
Furthermore, $\rho_s$  and $\Theta_s$ are the density and potential temperature 
of the reference state which is chosen to be isentropic (i.e., $\Theta_s={\rm const}$)
and in hydrostatic equilibrium;   ${\bm g}=GM/r^2 \bm{\hat{e}}_r$ 
is the gravity acceleration, $G$ and $M$ are the gravitational 
constant and the stellar mass, respectively, and $\mu_0$ is the magnetic 
permeability. The potential temperature, $\Theta$, is related to the specific
entropy: $s=c_p \ln\Theta+{\rm const}$. 
 
The term $\Theta'/\tau$ represents the balancing action of the turbulent 
Reynolds heat flux responsible for maintaining the steady axisymmetric solution 
of the stellar structure 
\citep[see section 1.2 and Annexe B in][for details]{Cossette15}.
In this work the 
parameters of the ambient state 
are slightly different from those in equation (7) of \cite{GSKM13b} \S3.1. 
For the models including a tachocline, here we use the polytropic indexes
$m_r=2$ and $m_{\rm cz}=1.499978$, together with the transition 
width $w_t=0.015\Rs$.  In the models without a tachocline, the 
polytropic index is 
constant $m_{\rm cz}=1.499985$. These
values result in convective motions with similar $\Ro$ for
both types of models. The relaxation
time of the potential temperature perturbation for all models 
is $\tau=1.036\times10^8$ s ($\sim 3.3$ yr).

The equations are solved numerically using the EULAG-MHD code  
\citep{GCS10,RCGS11,SC13,GSKM13b}, a spin-off of the hydrodynamical
model EULAG predominantly used in atmospheric and climate research
\citep{PSW08}. 
The time evolution is calculated using a special semi-implicit method based on a 
high-resolution, non-oscillatory forward-in-time advection scheme MPDATA 
\citep[Multidimensional Positive Definite Advection Transport Algorithm;][]{S06}. 
The truncation terms in MPDATA  evince
viscosity comparable to the explicit SGS viscosity used in large-eddy 
simulation (LES) models \citep{ES02,DXS03,MSW06}. Thus, the results of 
MPDATA are often interpreted as implicit LES or ILES \citep{GMR07}. 

For the velocity field we use impermeable, stress-free conditions at the 
top and bottom surfaces of the shell; whereas the magnetic field is
assumed to be radial at these boundaries. Finally, for the thermal
boundary condition we consider zero divergence of the convective flux 
at the bottom and zero flux at the top surface.

\section{Results}
\label{s.res}

\begin{table*}[ht]
\begin{center}
\caption{Simulation parameters and outputs: 
$\Ra^*=\frac{1}{\cp \Omega_0^2} g \frac{\partial s_e}{\partial r}$, here
$s_e$ is the specific entropy of the ambient state, 
$\urms$ is the volume averaged rms velocity (in m/s) in the unstable layer,
$\Ma=\urms/\cs^*$ is the Mach number, with $\cs^*=\sqrt{\gamma R T_s^*}\;|_{r=0.85\Rs}$
being the sound speed at the middle of the unstable layer,
$\Ro=\frac{\urms}{2\Omega_0 L}$, and $\chi_{\Omega} = (\Omega_{\rm{eq}}-
\Omega_{\rm {p}})/\Omega_0$, where
$\Omega_{\rm{eq}}=\mean{\Omega}(\Rs,0^{\circ})$, $\Omega_{\rm{p}}
=\mean{\Omega}(\Rs,60^{\circ})$ are the surface rotation rates at $0$
and $60$ degrees in latitude, and
$\Omega_0$ is the frame rotation rate expresed in terms of the 
solar rotation rate $\Omega_{\odot}$. The growth rate of the magnetic 
field, $\lambda$, is given in Tesla yr$^{-1}$. The kinetic and magnetic 
energy densities, in Jm$^{-3}$ are: $e_{\rm K}=\mean{\rho}_s \mean{\bm u}^2/2$,  
$e_{\rm M}=\mean{\bm{B}}^2/2\mu_0$, $e_{\phi}=\mean{B}_{\phi}^2/2\mu_0$ and
 $e_{p}=(\mean{B}_r^2+\mean{B}_{\theta}^2)/2\mu_0$. 
Finally, the full cycle period, $T_{\rm M}$, is expressed in years.
Models starting with the letter CZ consider the convection zone only, 
while models starting with RC include both, radiative and convective
zones. The number of grid mesh points is $N_r=47$, $N_{\theta}=64$  and
$N_{\phi}=128$ for CZ models, and $N_r=64$ for RC models. 
\label{tbl.1}}
\vspace{0.5cm}
\begin{adjustbox}{max width=\textwidth}
\begin{tabular}{cccccccccccc}
\tableline\tableline
Model &$\Omega_0$ &$\Ro$ & $\Ra^*$ & $\urms$ & $\Ma(10^{-4})$ & $\chi_{\Omega}$ &
      $\lambda$ & $e_{\rm{M}}/e_{\rm{K}}$ & $e_{\phi}/e_{\rm{K}}$ & 
      $e_{{p}}/e_{\rm{K}}$ & $T_{\rm{M}}$  \\
\tableline
CZ01 &$2\Omega_{\odot}$& 0.030 &1.29 & 55.12 & 3.85 &0.09 & 2.26 &0.098 &0.080 &0.018 &2.26   \\
CZ02 &$\Omega_{\odot}$&   0.067 &5.16 & 60.91 & 4.25 &0.17 & 0.65 &0.010 &0.009 &0.001 &2.21   \\
CZ03 &$\Omega_{\odot}/2$&  0.150 &20.65 & 67.18 & 4.69&0.18 & 0.56 &$2\cdot10^{-4}$ &$2\cdot10^{-4}$ &$1\cdot10^{-5}$ &-      \\
\tableline
RC01 &$2\Omega_{\odot}$&0.033 &1.36    &60.80    &4.10    & 0.07& 0.84  &0.249 &0.042 &0.206 &-\\
RC02 &$\Omega_{\odot}$&0.069   &5.45    &62.27    &4.30    & 0.05& 0.85  &0.184 &0.163 &0.020 &34.5\\
RC03 &$\Omega_{\odot}/2$&0.161  &21.80   &72.88    &5.04    & 0.28& 0.06  &0.004 &0.004 &$3\cdot10^{-4}$  &-\\
\tableline
\tableline
\end{tabular}
\end{adjustbox}
\end{center}
\end{table*}

We have performed two sets of simulations: 1) only for the unstable 
stratified convection zone (CZ models), and  2) for the convection zone with 
a convectively stable radiative zone at the bottom of the domain (RC models). 
The first set does not support the formation of a strong radial shear at the base 
of the convection zone (the tachocline), and thus excludes it as a source of 
magnetic field.
The second  set naturally leads to the  development of a tachocline
and therefore to the $(\bm{B}\cdot \nabla)\Omega$ source of the toroidal
field acting in the
tachocline. The input and resulting parameters of the models for three 
different rotation rates, $\Omega_0$ (corresponding to different Rossby
numbers)\footnote{Since we change the value of the $\Ro$  only 
by changing the value of the frame rotation rate, use both terms 
interchangeably to express the dependence of the results 
with $\Omega_0$.} are summarized in Table \ref{tbl.1}. 
\vfill\eject

\subsection{Dynamo models without tachocline}
\label{s.tac}
\begin{figure}[H]
\begin{center}
\includegraphics[width=0.42\columnwidth]{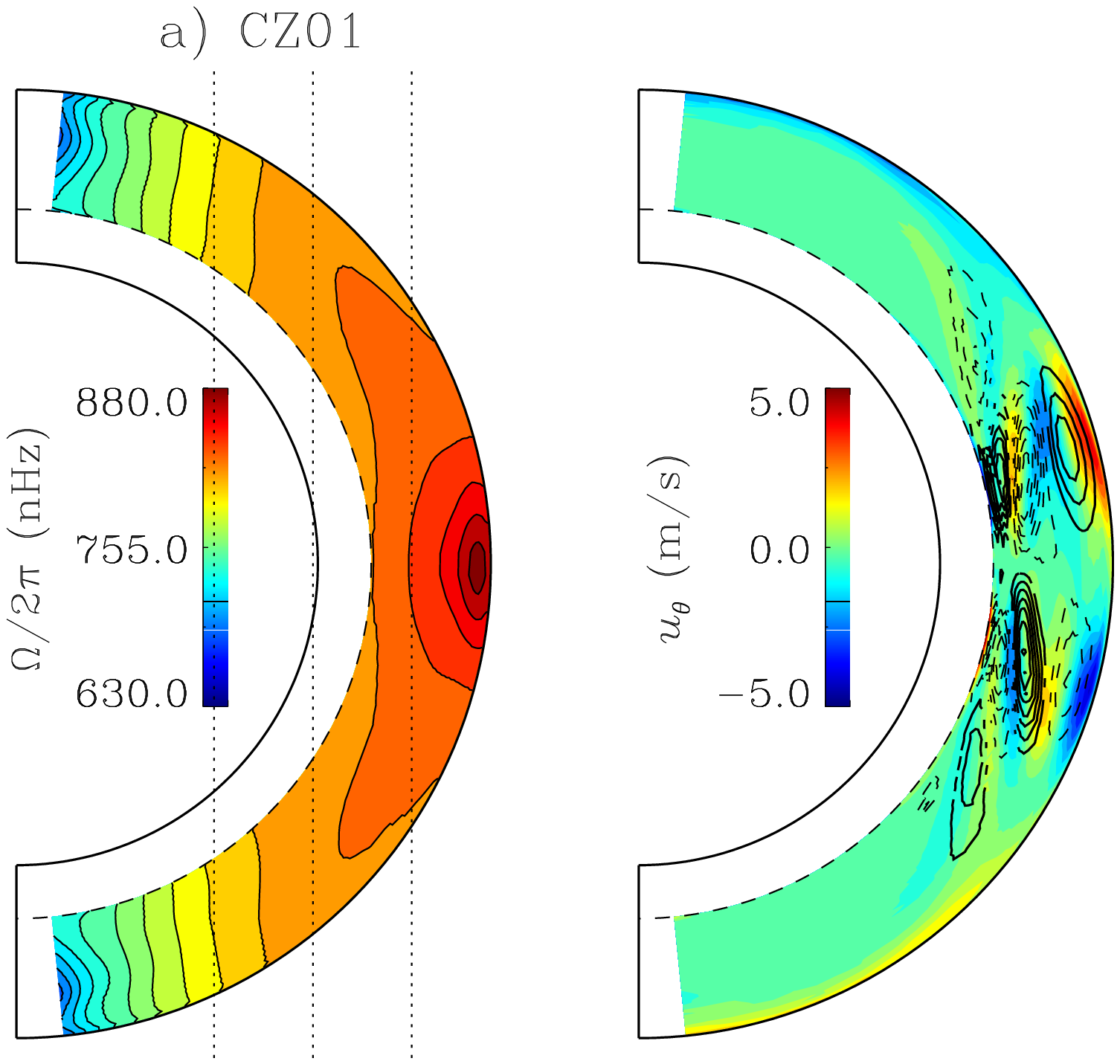}
\includegraphics[width=0.57\columnwidth]{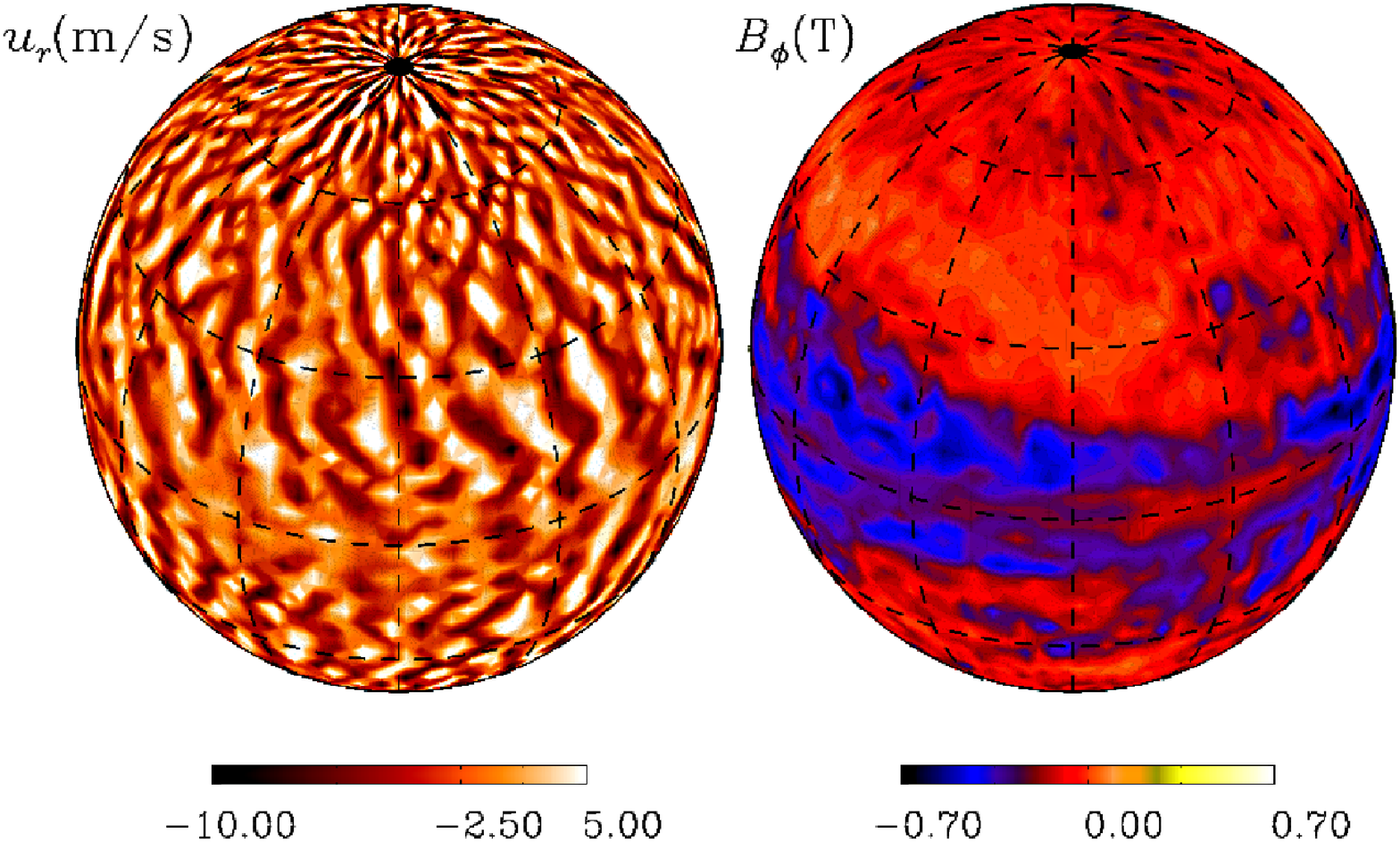}\\
\includegraphics[width=0.42\columnwidth]{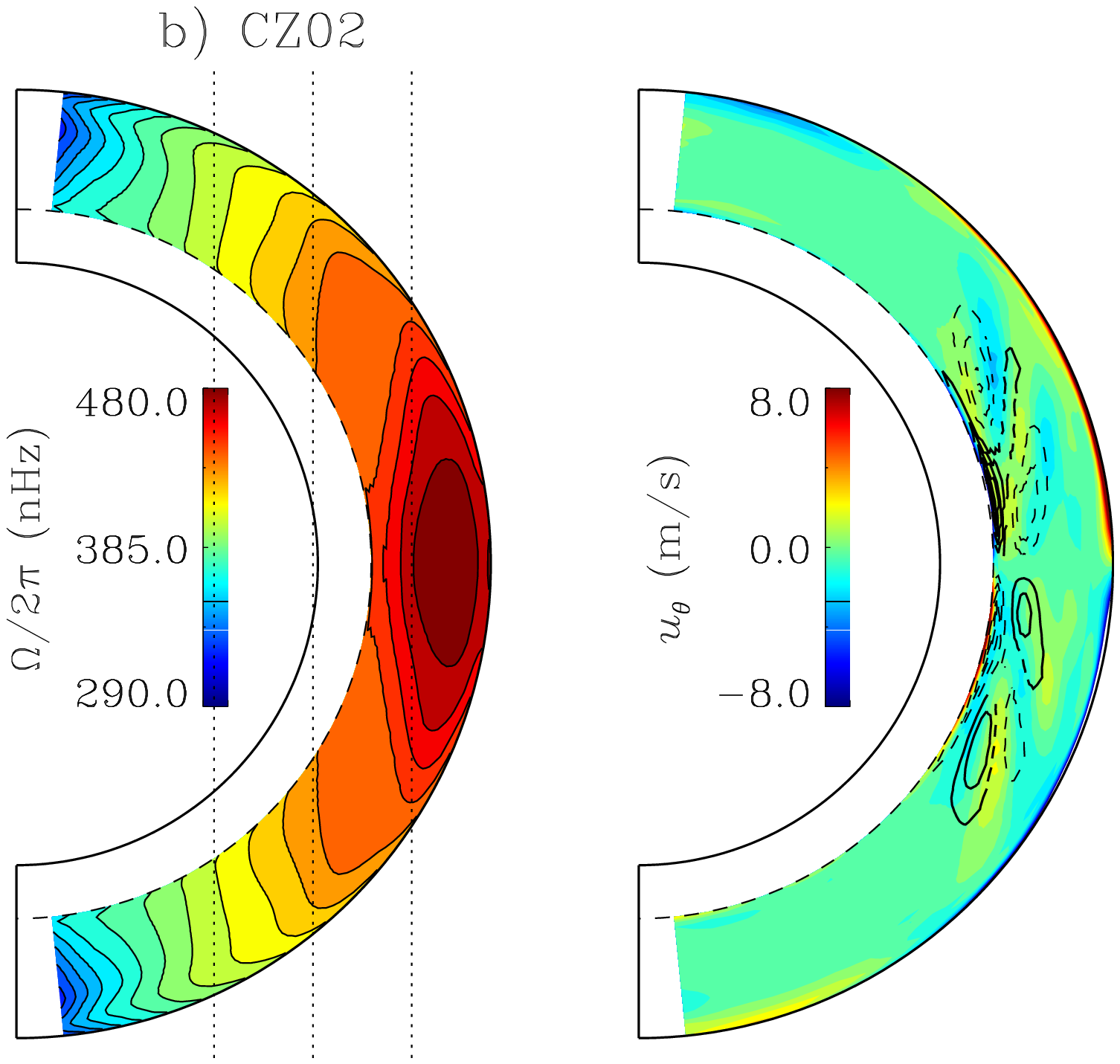}
\includegraphics[width=0.57\columnwidth]{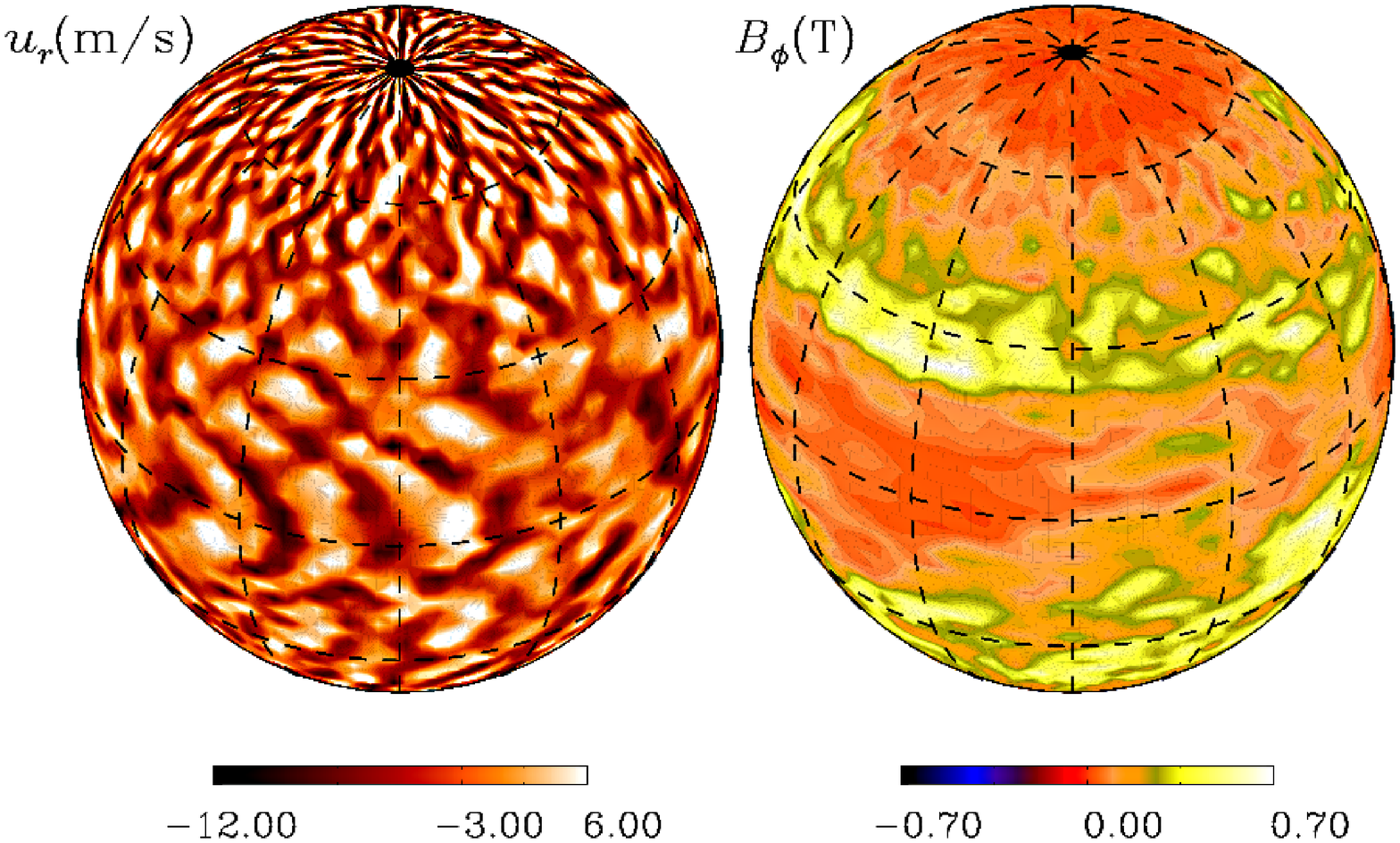}\\
\includegraphics[width=0.42\columnwidth]{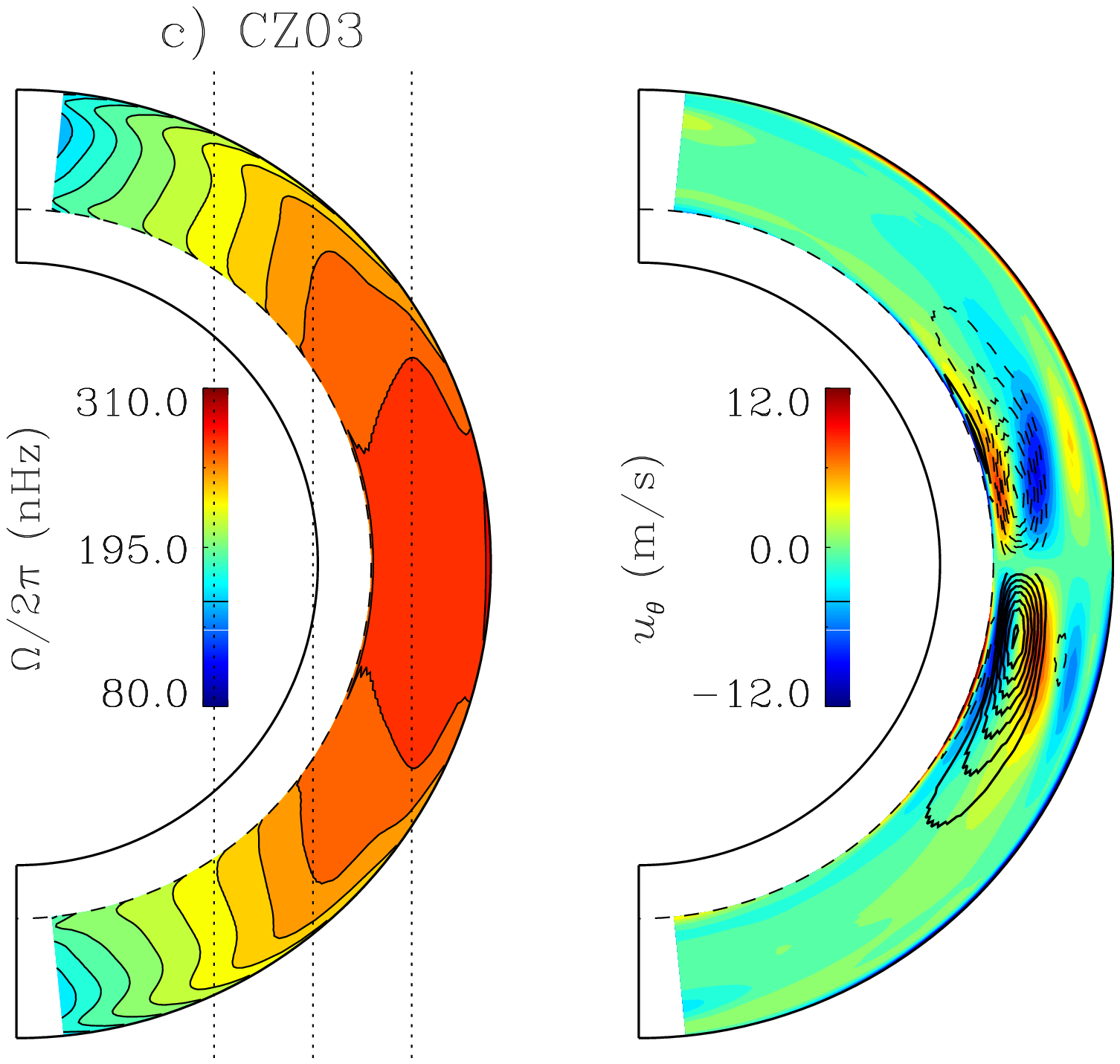}
\includegraphics[width=0.57\columnwidth]{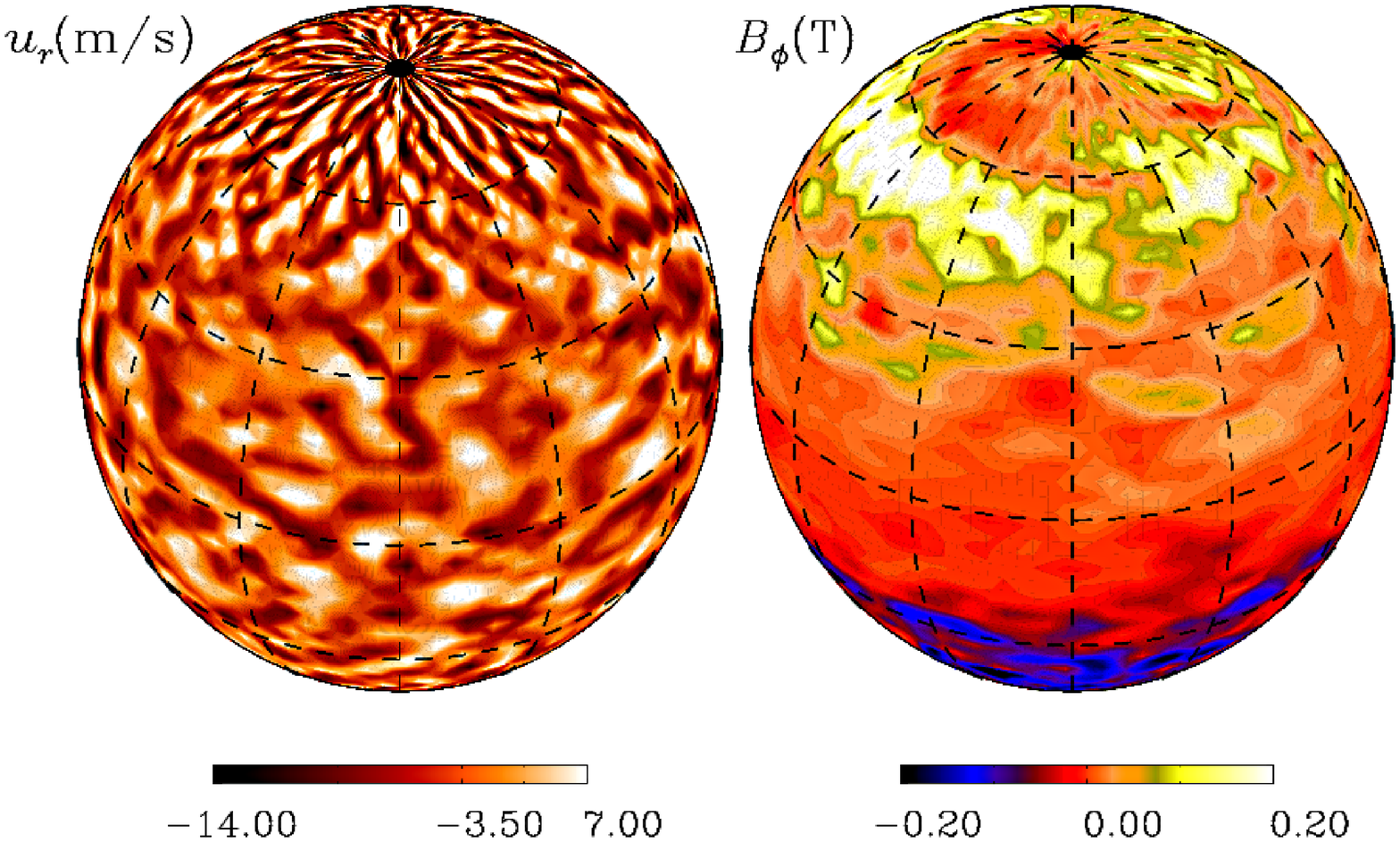}\\
\caption{From left to right: mean profiles of the rotation rate and meridional circulation, 
and snapshots at $r=0.95\Rs$ of the vertical velocity, $u_r$, and the toroidal field, 
$B_{\phi}$, for the simulated models a) ~CZ01, b) ~CZ02, and c) ~CZ03. In the meridional circulation panels
continuous (dashed) lines correspond to clockwise (counterclockwise) circulation.
The background filled contours show the latitudinal velocity, $\mean{u}_{\theta}$.
The profiles of differential rotation and meridional circulation  
correspond to mean azimuthal values averaged over $\sim3$
years during the steady state phase of the simulation.}
\label{fig.df1}
\end{center}
\end{figure}

In Figure \ref{fig.df1} we present, from left to right, the meridional 
profiles of the differential rotation, meridional circulation, and snapshots 
of the vertical velocity, $u_r$, and the toroidal magnetic field $B_{\phi}$ resulting
from the models a) CZ01, b) CZ02 and c) CZ03. The differential rotation and 
meridional circulation correspond to averages over longitude and time ($\sim 3$ years).
The instantaneous orthographic projections of $u_r$ and $B_{\phi}$ allow us to 
distinguish the character of the convective flow and the distribution of the
toroidal field for each model. 
Unlike our previous purely hydrodynamic simulations, in which we found
a  critical $\Ro$ value dividing the rotation profiles into solar-like and 
anti-solar rotation types \citep{GSKM13b}, in the present simulations with
magnetic field all the  models 
exhibit the solar-like differential rotation, irrespective of the Rossby number. 
This property is related to the influence of the dynamo-generated 
magnetic field on the fluid dynamics. 
\cite{KKBOP15} have found that for the MHD case
the transition occurs at higher values of $\Ro$. 
For instance, model ~CZ01 shows iso-rotation contours mainly aligned 
along the rotation axis (see Fig. \ref{fig.df1}a).  In  models ~CZ02 
(Fig. \ref{fig.df1}b)   and CZ03 (Fig. \ref{fig.df1}c)
the rotation profile exhibits  clear conical-shape
contours that resemble the rotation of the solar convection zone 
inferred by helioseismology \citep{SCHOU98}.  
During early stages of the models evolution, when the magnetic field
influence is negligible, the rotation contours are aligned along the
rotation axis (the so-called Taylor-Proudman balance), and then
transformed into the conical-shape contours. This
indicates that both Reynolds and Maxwell stresses contribute to 
the distribution of angular momentum.  Later, it is remarkable that even
model ~CZ03, in which the final magnetic energy is $10^4$ times
smaller than the kinetic energy, the rotation law departs from 
the Taylor-Proudman balance. A detailed discussion of the angular 
momentum balance is out of the scope of this paper and will be 
addressed in a subsequent paper. 

Another remarkable feature of models ~CZ02 and ~CZ03 is
the natural development of a well-defined NSSL.  The radial shear 
in this layer is negative and extends from the equator to the poles 
as observed in the Sun. In agreement with the results found  
in \cite{GSKM13b},  the NSSL arises  from the appropriate choice of the 
ambient state, $\Theta_e$, and the relaxation time of the
potential temperature perturbations, $\tau$. 
As mentioned above, these quantities differ from the ones used 
in \cite{GSKM13b} in such a way that $\Ro$ becomes larger at the top 
of the domain (i.e., in a thin layer close to the surface where the 
buoyancy force becomes much stronger than the Coriolis force). Although 
the number of mesh points is not sufficient to resolve super-granulation 
scales which may be important for the formation of the 
solar NSSL, our model succeeded  in reproducing  similar effects  
in the resolved scales.
Furthermore, the latitudinal angular 
velocity gradient in both models corresponds well to the 
solar rotation, i.e., $\chi_{\Omega} = (\Omega_{\rm{eq}}-
\Omega_{\rm {p}})/\Omega_0 = 0.18$, where
$\Omega_{\rm{eq}}$ is the equatorial angular velocity, $\Omega_{\rm{p}}$
is the angular velocity at $60^{\circ}$ latitude (both quantities are 
computed from the temporally and azimuthally averaged profile 
of $\Omega$), and 
$\Omega_0$ is the angular velocity of the reference frame. 

As far as the meridional circulation is concerned, models ~CZ01 and ~CZ02 show
a multicellular pattern (second column of Fig. \ref{fig.df1}a and b), 
while model ~CZ03 shows a dominant (counter) 
clockwise cell in the (Northern) Southern hemisphere with a latitudinal
velocity, $u_{\theta}\sim 12$ m s$^{-1}$ (about twice the value of 
$u_{\theta}$ in models ~CZ01 and ~CZ02). This cell is located
near the base of the convection zone. Another cell, with a smaller 
velocity and opposite circulation direction appears in the 
subsurface  layer (second column of Fig. \ref{fig.df1}c). 

\begin{figure}[htp]
\begin{center}
\includegraphics[width=0.49\columnwidth]{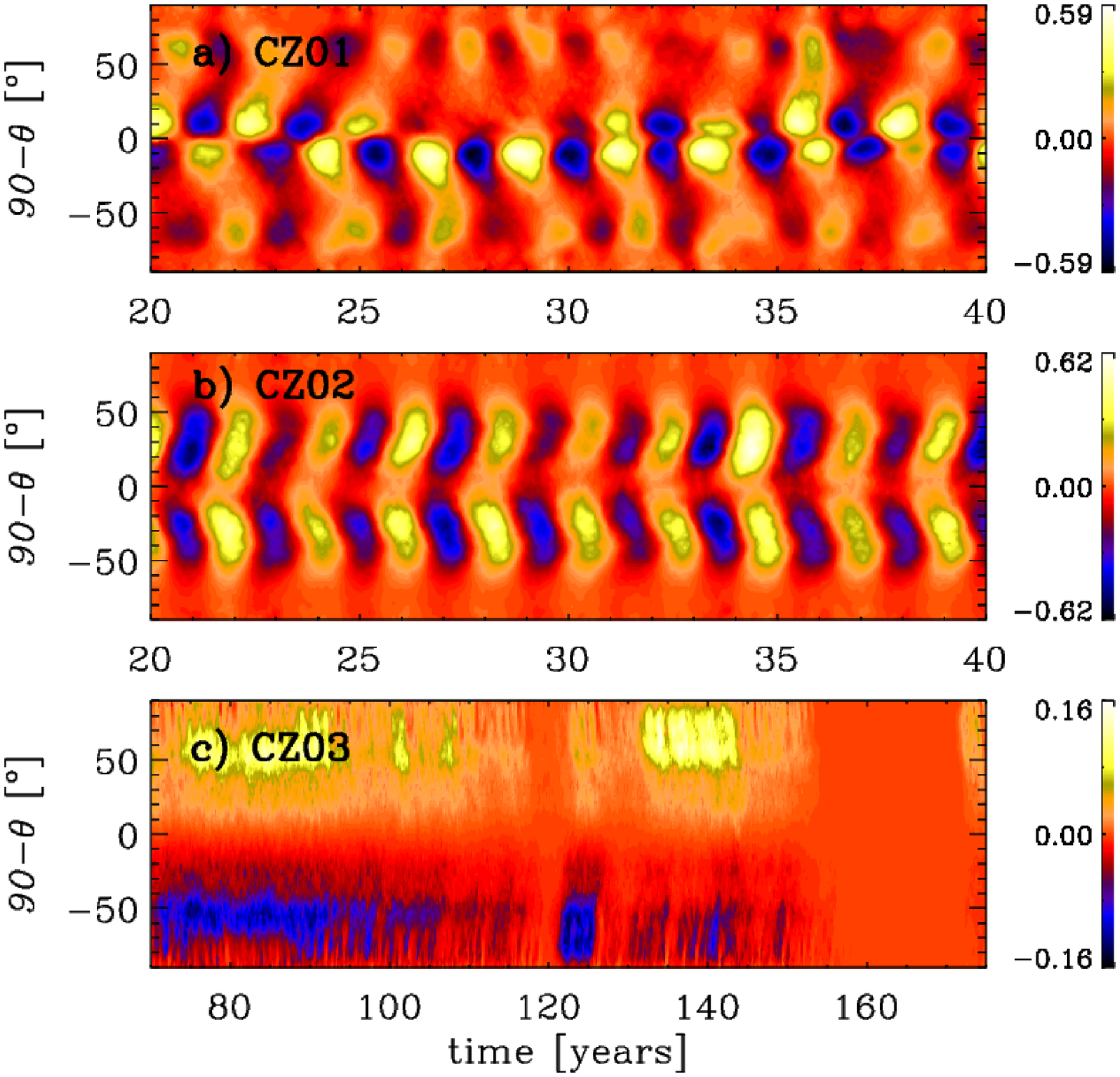}
\includegraphics[width=0.49\columnwidth]{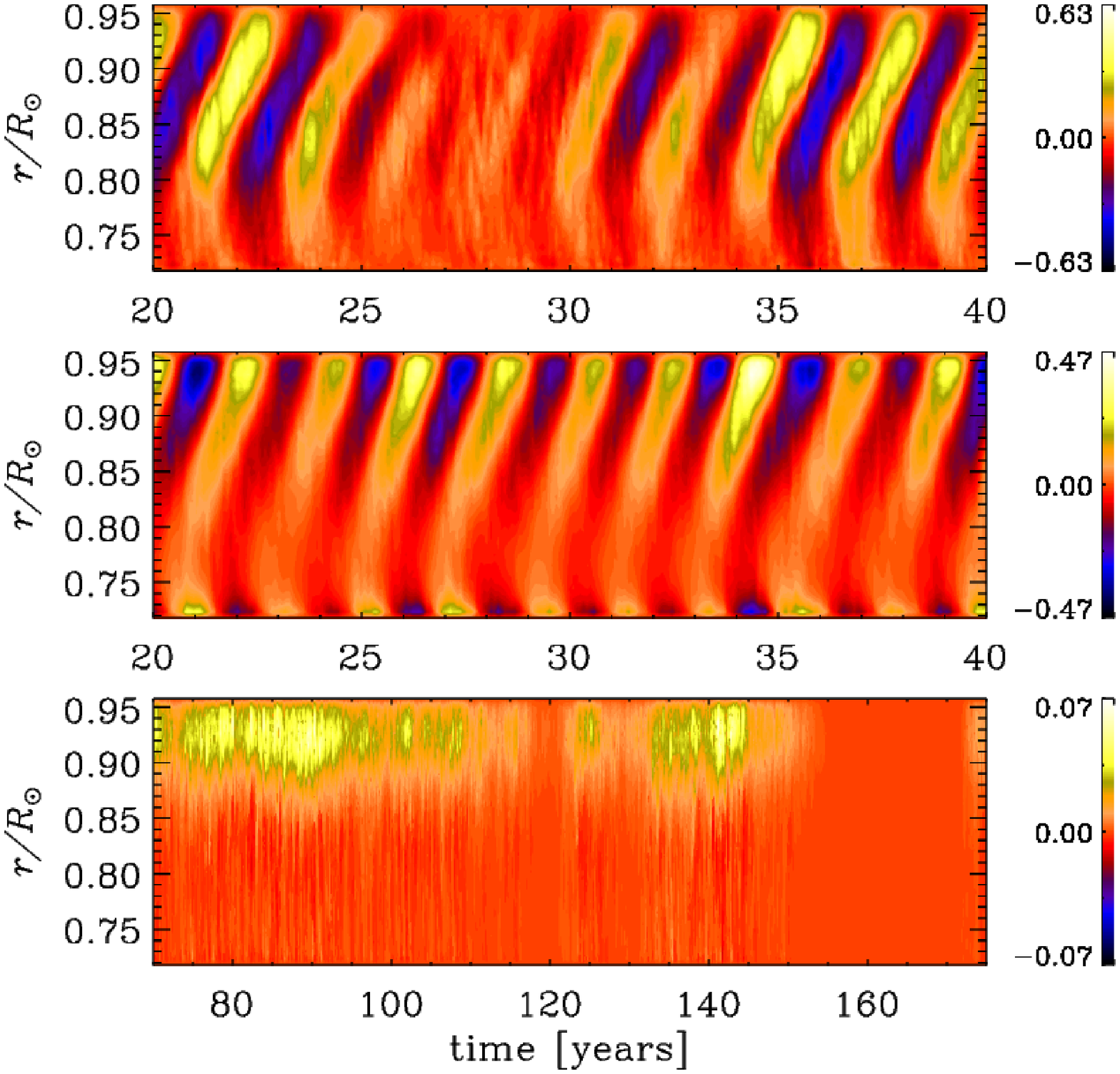}
\caption{Time-latitude (left) and radius-latitude (right)  
diagrams of the toroidal magnetic field, $\mean{B}_{\phi}$,  
for models a) ~CZ01, b) ~CZ02 and c) ~CZ03. On the left, 
the contours show $\mean{B}_{\phi}$, in Tesla, 
at $r=0.95\Rs$. On the right, the contours are taken at 
$30^{\circ}$ latitude. Only a fraction of time of the simulated 
statistically steady state is shown in each plot.}
\label{fig.bd1}
\end{center}
\end{figure}

An oscillatory large-scale dynamo action is observed in models 
~CZ01 and ~CZ02 as can be seen in  Fig. \ref{fig.bd1}a,b. 
In both cases the toroidal magnetic field is generally symmetric 
across the equator and reverses with a cycle period of $\sim2$
years (Table \ref{tbl.1}).  For the model CZ02 we found a 
secondary $\sim6$-years cycle modulating the amplitude of the 
magnetic field.  These properties are at odd with the 
solar 22-year cycle but in agreement with other global models 
without tachocline \citep[e.g.,][]{ABMT15,WKKB14}.  In the convection zone the 
evolution of the magnetic fields is consistent with a pattern of 
dynamo waves propagating from the bottom to the top of the domain 
(Fig. \ref{fig.bd1}a,b). At the surface the field forms 
a $\pm 50^{\circ}$ belt around the equator. 
The latitudinal migration of the field is slightly poleward. 
The long term evolution of the simulations ~CZ01 and ~CZ02 indicates 
extended periods of minimal or maximal activity (see for instance
the extended minimum in the Northern hemisphere of the model CZ01, 
Fig. \ref{fig.bd1}a, between 20 and 30 years). The 
amplitude of the dynamo-generated magnetic field depends on the
rotation rate, reaching $\sim 10\%$ and $\sim 1\%$ of the kinetic
energy of the system, for the models ~CZ01 and ~CZ02, respectively 
(see Table \ref{tbl.1}). In model ~CZ03 (Fig. \ref{fig.bd1}c) the 
magnetic field grows exponentially but saturates at $\sim 0.02\%$
of the kinetic energy, (see Table \ref{tbl.1}). The 
magnetic field in this case is nearly steady. Its time evolution
shows periods when the field amplitude abruptly decays, 
stays in a minimum state for some years, and then quickly grows again 
to its saturation amplitude without polarity reversals (Fig. \ref{fig.bd1}c). 
As indicated in Table \ref{tbl.1}, the growth rate of the dynamo 
is proportional to the rotation rate.

\begin{figure}[H]
\begin{center}
\includegraphics[width=0.7\columnwidth]{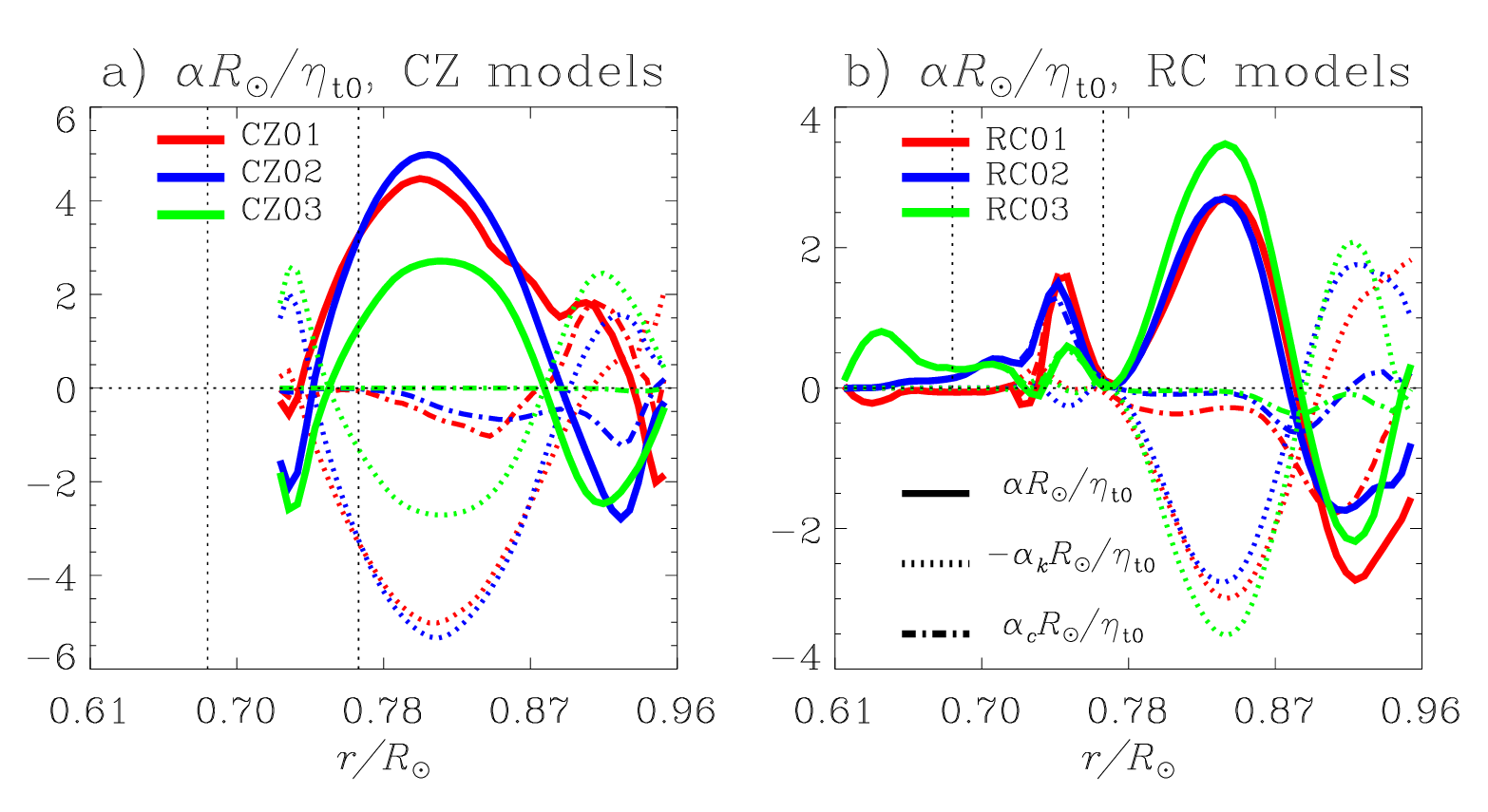}\\
\includegraphics[width=0.7\columnwidth]{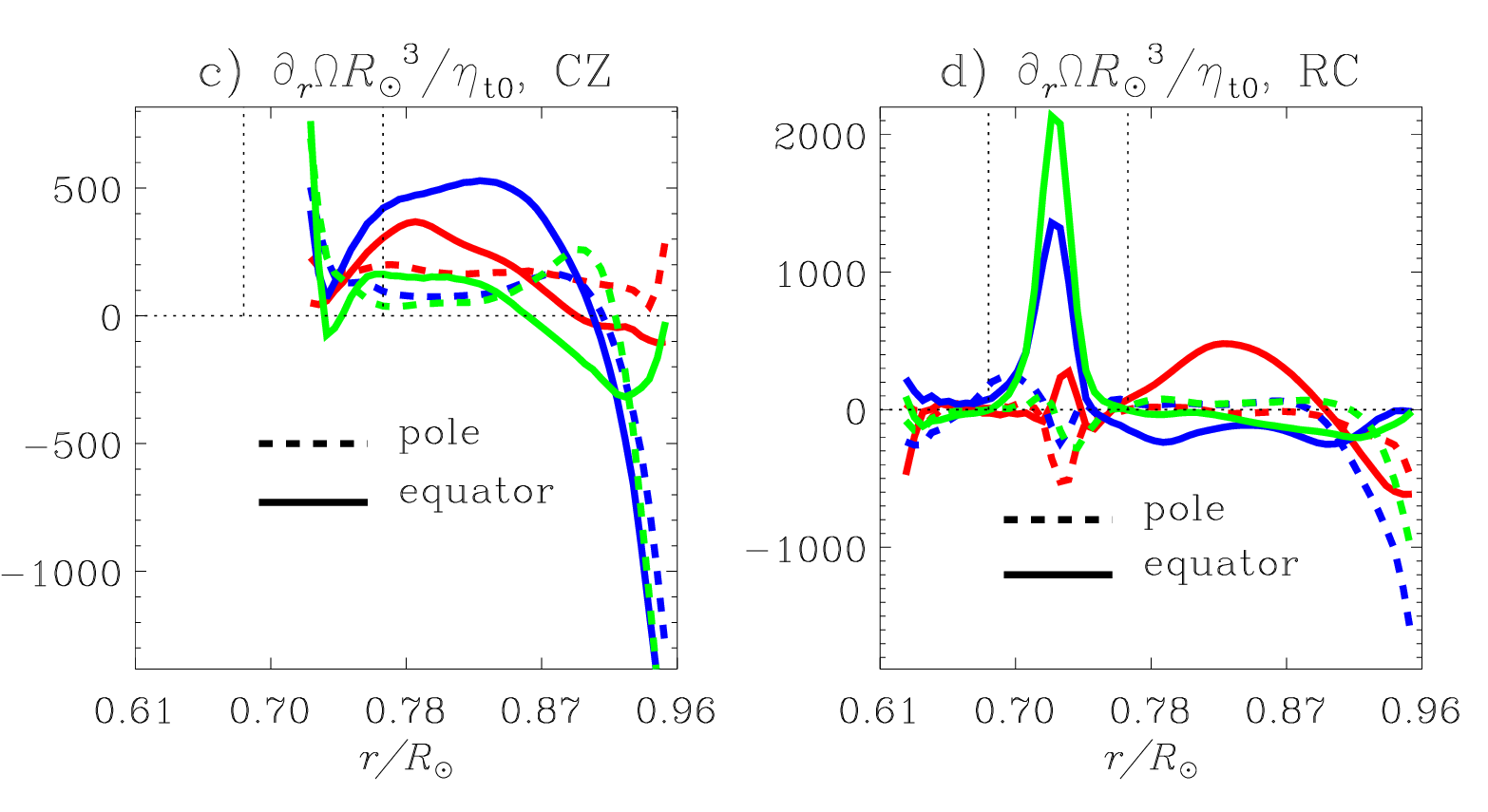}\\
\includegraphics[width=0.7\columnwidth]{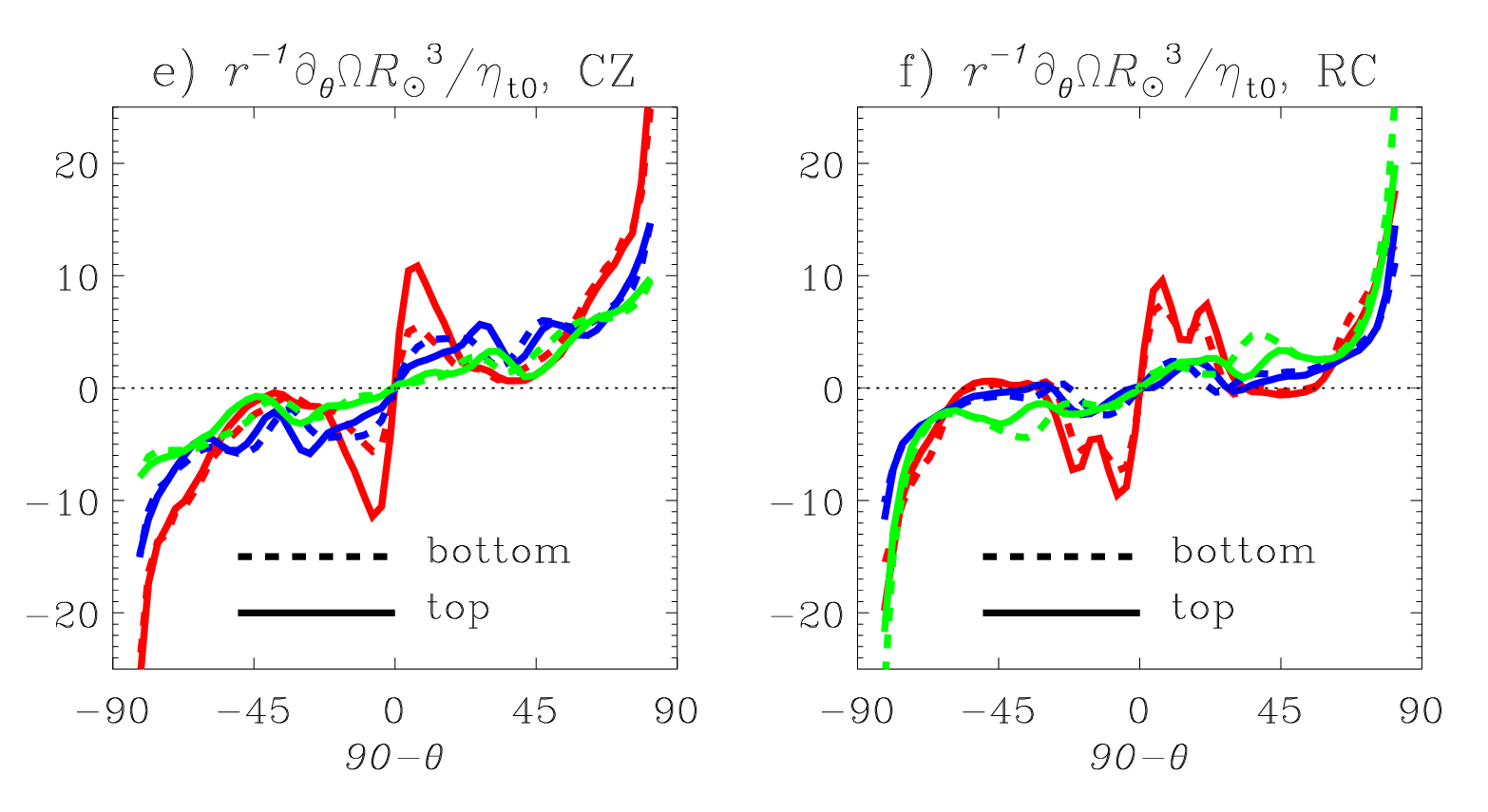}
\caption{\small{Magnetic field source terms (see Eqs. 8-9) 
for models CZ (left panels), and RC (right). Panels a) and b) 
show the FOSA estimation for 
$\alpha \Rs /\etato$. Dotted and dot-dashed lines represent 
the kinetic and magnetic contribution, respectively. The continuous line 
is the resultant, non-dimensional, total $\alpha$-effect. The profiles 
of $\alpha$ are latitudinal averages in the northern hemisphere.
Middle and bottom rows show $\partial_r\Omega \Rs^3/\etato$ 
and $r^{-1}\partial_{\theta} \Omega \Rs^3/\etato$, respectively. 
In panels c) and d) continuous (dashed) lines indicate latitudinal 
averages at lower, $20^{\circ}-40^{\circ}$, (higher, 
$60^{\circ}-80^{\circ}$), latitudes.  (Note that in panel d) the
value of $\partial_r \Omega \Rs^2$ has been divided by $2$.) 
In panels e) and f) continuous (dashed) lines are vertical averages 
in the top, 
$0.9\Rs \le r \le 0.95\Rs$ (bottom, $0.72\Rs \le r \le 0.77\Rs$), of
the domain.  $\etato$ is a radial average of $\etat$, defined in 
Eq. (\ref{eq.etat}), for $r \ge 0.72 \Rs$.}
\label{fig.alpha-omega}}
\end{center}
\end{figure}

\begin{table}[htb]
\begin{center}
\caption{Dynamo coefficients computed from the simulations results. The dynamo numbers
are computed with Eqs. (\ref{eq.dc}) and (\ref{eq.etat}). The $0$ index 
in $C_{\alpha}$ and $C_{\Omega}$ refers to the maximum value of each quantity, in $\etat$,
to its average over the CZ (in ${\rm m}^2 {\rm s}^{-1}$). 
The dynamo numbers $D_r=C_{\alpha 0}C^r_{\Omega0}$ and 
$D_{\theta}=C_{\alpha 0}C^{\theta}_{\Omega0}$, provide an indication of the dynamo 
efficiency when considering the radial and the latitudinal shear, respectively. Finally 
max($\mean{B}_{\phi}$), max($\mean{B}_{\theta}$) and max($\mean{B}_{r}$) are the maximum
absolute values of the large-scale components of the field at $r=0.95\Rs$ (the values in
brackets are measured at $r=0.72\Rs$) in units of Tesla.  
\label{tbl.2}}
\vspace{0.5cm}
\begin{adjustbox}{max width=\textwidth}
\begin{tabular}{cccccccccccc}
\tableline\tableline
Model & $\etato(10^9)$ & $C^k_{\alpha 0}$ & $C^m_{\alpha 0}$ & $C_{\alpha 0}$ & 
$C^r_{\Omega 0}(10^3)$ &$C^{\theta}_{\Omega 0}$ & $D_r$ & $D_{\theta}$ & 
max($\mean{B}_{\phi}$) & max($\mean{B}_{\theta}$) &max($\mean{B}_{r}$)\\
\tableline
CZ01 & 0.99 & 5.01 & 1.84  & 4.47  & 0.37 & 27.28 & 1.65 & 122.90 & 0.60 &0.18 & 0.03  \\
CZ02 & 1.09 & 5.38 & 1.28  & 5.00  & 1.62 & 14.75 & 8.10 & 73.75  & 0.61 &0.13 & 0.06\\
CZ03 & 1.28 & 2.72 & 0.08  & 2,72  & 1.69 & 9.71  & 4.59  & 26.41  & 0.15 & 0.05 &0.02  \\
\tableline
RC01 & 1.08 & 3.03 & 2.06  & 2.75  & 0.81 & 19.65 & 2.22  & 54.03 & 0.52(0.19) & 0.10(0.80) & 0.04(0.26) \\
RC02 & 1.12 & 2.80 & 1.52  & 2.72  & 1.67 & 13.33 & 4.54 & 36.25 & 0.22(0.94) & 0.07(0.28) & 0.03(0.24) \\
RC03 & 1.25 & 3.56 & 0.64  & 3.52  & 4.98 & 21.51 & 17.53 & 75.71 & 0.20(0.71) & 0.04(0.37) & 0.03(0.20) \\
\tableline
\tableline
\end{tabular}
\end{adjustbox}
\end{center}
\end{table}

The spatio-temporal evolution of the magnetic field in the models 
can be interpreted in terms of its main sources according to 
the mean-field dynamo theory, i.e., the $\alpha$ and $\Omega$ effects.
 The profile of $\alpha$ is estimated
by using the first-order smoothing  approximation 
\citep[FOSA, see for details][]{BS05} as follows:
\begin{equation}
\alpha=\alpha_k + \alpha_m = -\frac{\tau_c}{3}h_k + \frac{\tau_c}{3} h_c
\end{equation}
where $\tau_c=H_{\rho}/\urms$ is 
the turnover time of the convection, $H_{\rho}^{-1}=d{\rm ln}\rho/dr$ 
is the density length scale, and 
\begin{eqnarray}
h_k&=&\mean{{\bm u}'\cdot \nabla \times {\bm u}'}\\\nonumber
	h_c&=&\frac{\mean{{\bm b}'\cdot \nabla \times {\bm b}'}}{\mu_0 \rho_s}
\end{eqnarray}
are the small-scale (note the prime over ${\bm u}$ and ${\bm b}$) kinetic
and current helicities, respectively (the overbars denote azimuthal and 
temporal averages over 3 years). 
As it is common in the mean-field dynamo theory,  we present the source terms
them as ratios between the induction and diffusion times: 
\begin{eqnarray}
\label{eq.dc}
C_{\alpha}^k= \frac{\alpha_k \Rs}{\etato} \; &,& 
C_{\alpha}^m= \frac{\alpha_m \Rs}{\etato} \;, \\\nonumber
C_{\Omega}^r= \frac{\partial_r \Omega \Rs^3}{\etato} \; &,& 
C_{\Omega}^{\theta}= \frac{\partial_{\theta} \Omega \Rs^3}{r \etato} \;,
\end{eqnarray}
where  $\etato$ is the FOSA estimation of the turbulent 
magnetic diffusivity,
\begin{equation}
\label{eq.etat}
\eta_{t0}=\frac{1}{3}\tau_c \urms^2 \;,
\end{equation}
averaged over the entire convection zone, $0.72\Rs \le r \le 0.96\Rs$ 
(the values of $\eta_{t0}$ are presented in Table \ref{tbl.2}).

In the top, middle and bottom rows of 
Fig. \ref{fig.alpha-omega}  we present the normalized profiles of 
$\alpha$, $\partial_r \Omega$ and 
$r^{-1}\partial_{\theta} \Omega$, respectively,  computed 
from the simulation results. Fig. \ref{fig.alpha-omega}a shows 
the radial profiles of the kinetic 
and magnetic contributions to $\alpha$ presented by dotted and 
dot-dashed lines, respectively,  while the resultant $\alpha$ is shown with continuous 
lines.  The vertical profiles of the radial shear, Fig. \ref{fig.alpha-omega}c,
are shown at two different 
ranges of latitude: a higher latitude profile (dashed lines) was 
computed averaging over $60^{\circ}$ and $80^{\circ}$ latitude,
and the lower latitude profile (continuous line) was averaged between 
$20^{\circ}$ and $40^{\circ}$ latitude. Finally, the latitudinal shear was also 
computed for two different ranges of depths, Fig. \ref{fig.alpha-omega}e . 
The bottom profile (dashed line) 
corresponds to an average between $0.72\Rs \le r \le 0.77\Rs$, while the 
top profile (continuous line) is an average between $0.90\Rs \le r \le 0.95\Rs$.
Presenting these quantities as radial profiles, instead of contours in the 
meridional plane, allows to straightforwardly compare the different dynamo models.
The red, blue and green lines correspond to models ~01, ~02 and ~03, respectively. 
Furthermore, Figure \ref{fig.alpha-omega} allows us to directly 
compare the CZ and RC models (to be discussed later), shown in the left 
and right panels, respectively. 

Before describing the evolution of the magnetic fields, it is 
instructive to discuss how the source terms change as a function of the 
Rossby number. Table \ref{tbl.2} presents the maximal absolute 
values of the quantities depicted in Fig. \ref{fig.alpha-omega} (denoted
with a $0$ index), as well as 
the maximal amplitude of the mean magnetic field components at the surface 
level. (Some values 
at the boundaries can be very large without really contributing
to the dynamo, so we have disregarded them.)
In general, there are no major differences between the profiles of the
kinetic $\alpha$-term, $C^k_{\alpha}$. In all the cases it is positive in the
bulk of the convection zone and changes sign near the boundaries.
The radius at which this change of sign occurs is shallower (deeper) 
for the faster (slower) rotating model.  
There is not a clear relation between $C^k_{\alpha 0}$ and $\Ro$, it
is only noticeable, however, that the faster models, ~CZ01 and ~CZ02, 
have larger kinetic helicity than the slower model, ~CZ03. 
As for the the magnetic $\alpha$-effect,
we should first notice that its amplitude is comparable 
to the kinetic $\alpha$. However, $C^m_{\alpha 0}$  is  
inversely proportional to $\Ro$.  This is expected since the models 
with faster rotation develop stronger magnetic fields. Correspondingly,
the total $\alpha$-effect is inversely proportional to the Rossby 
number.  For the models ~CZ01 and ~CZ02 the signs of $C^k_{\alpha}$  
and $C^m_{\alpha}$ in the middle of the convection zone are opposite. 
At the top 
of the domain the two terms have the same sign, and both contribute positively 
to the generation of magnetic field. In model ~CZ03 the contribution
of $C^m_{\alpha}$ is negligible, thus, only the kinetic helicity
contributes to the field generation. 

Regarding the rotational shear terms, we notice that for larger 
values of $\Ro$, the surface radial shear is stronger and the latitudinal 
shear is weaker. 
Note that the latitudinal profile of $C_{\Omega}^{\theta}$ 
(Fig. \ref{fig.alpha-omega}e)	
for model ~CZ01 shows a strong shear at the equator and the poles. 
This shear is reduced in models ~CZ02 and ~CZ03 as 
the vertical motions become more important. 
These fast vertical flows, in turn, are less affected by the
Coriolis force, resulting in a strong radial shear in the NSSL. 
The surface values of the magnetic field are also
inversely proportional to $\Ro$, possibly indicating that their
main source is the latitudinal shear. 

\begin{figure}[H]
\begin{center}
\includegraphics[width=0.69\columnwidth]{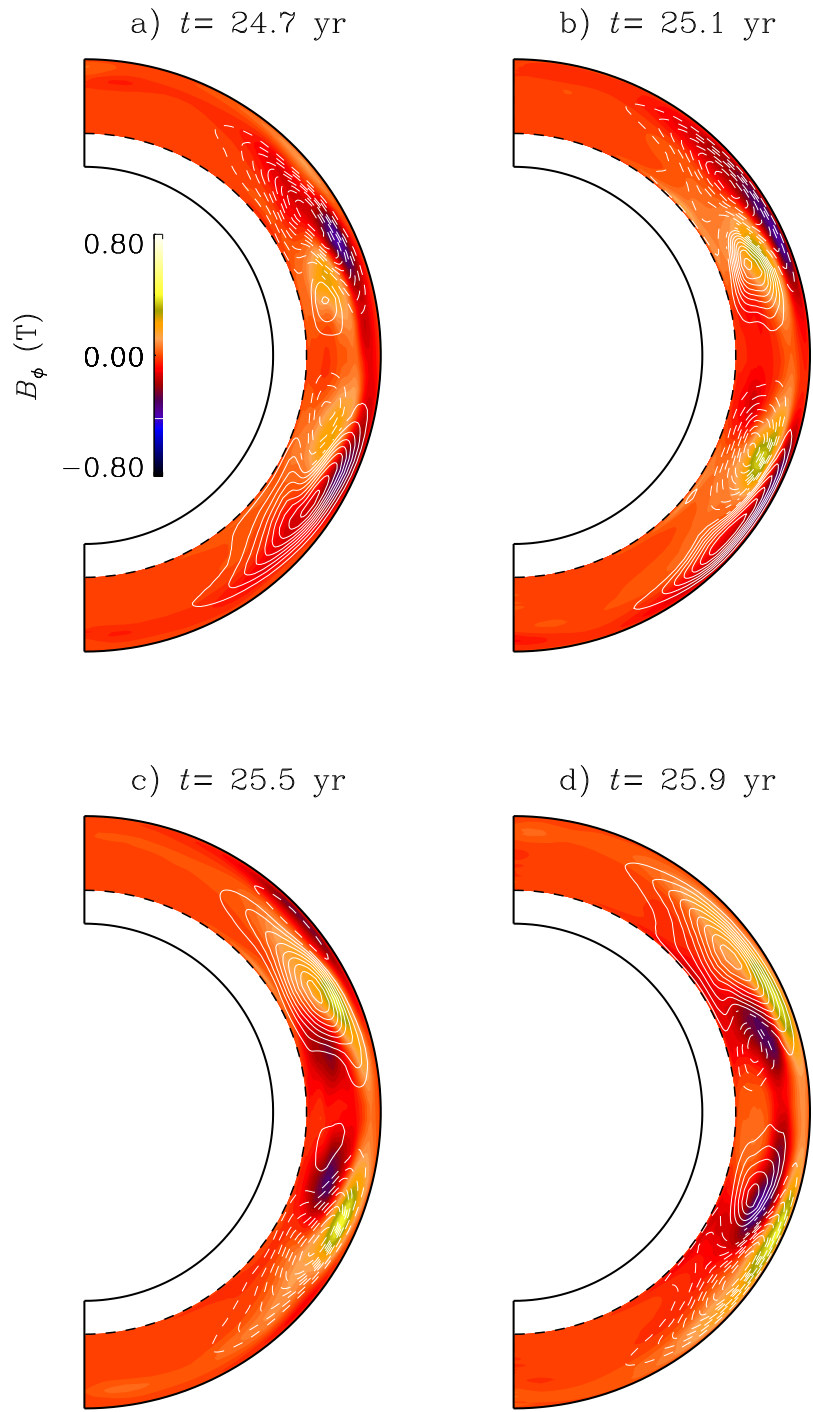}
\caption{Snapshots of the mean magnetic fields, $\mean{B}_{\phi}$ 
(color images), and poloidal magnetic field (contour lines) for the model CZ02. 
Continuous (dashed)
lines represent clockwise (counterclockwise) magnetic field direction.
The time series covers one polarity reversal (half dynamo cycle).}
\label{fig.revnt}
\end{center}
\end{figure}

The evolution of models ~CZ01 and ~CZ02, presented in the butterfly
diagrams of Fig. \ref{fig.bd1}a,b, is similar. 
They exhibit branches of magnetic field migrating in latitude from
the equator towards the poles. This pattern agrees with the  
Parker-Yoshimura sign rule, Eq. (\ref{eq.py}), for positive values of 
$\alpha \partial_r \Omega$ in almost the entire convection zone (see 
continuous red and blue lines in Fig. \ref{fig.alpha-omega}a, c). 
In the vertical direction the propagation of the magnetic field in
the middle of the CZ also agrees
with Eq. (\ref{eq.py}) for $\alpha \partial_{\theta} \Omega>0$. 
Because of the large values of $C^{\theta}_{\Omega}$ in model 
~CZ01, the branches of toroidal field show the large magnetic field
strength from the bottom of the convection zone to the top. In  model 
~CZ02, with a smooth 
latitudinal shear the stronger vertical branches of $\mean{B}_{\phi}$ 
start at a depth $\sim0.85\Rs$ (compare the right panels of Fig. 
\ref{fig.bd1}a and b).  At the upper radial levels of model ~CZ02,
the term $\alpha \partial_{\theta} \Omega$ becomes negative, and this should
change the direction of migration. However, at the latitude depicted
in Fig. \ref{fig.bd1}, only a slight change in the tilt of the 
branches is evident.  The morphology of the azimuthally averaged 
magnetic field of model ~CZ02, is depicted in Fig. 
\ref{fig.revnt}a-d. The filled contours show the positive (yellow) 
and negative (blue) toroidal field 
strength. The continuous (dashed) lines represent the clockwise 
(counterclockwise) poloidal field lines.  The four snapshots cover
one magnetic field reversal. In the same plot it can be observed
that the poloidal field follows a wave-like evolution
pattern similar to $\mean{B}_{\phi}$. The pattern starts 
to develop at the 
bottom of the domain at lower latitudes.  As the cycle evolves 
it grows and expands over the convection zone with the more 
intense regions (the innermost contours) migrating poleward and 
upward. In this case, 
however, the field is  antisymmetric with respect to the equator. 

\subsection{Dynamo models with tachocline}
\label{s.tac}

In this section we present three models (RC01, ~RC02 and ~RC03) 
with the tachocline.  All of them have the same stratification and 
differ only by the rotation rate, $\Omega_0$.
For all these cases, the magnetic field
evolves differently than in the ~CZ01, CZ02 and CZ03 models.

\begin{figure}[H]
\begin{center}
\includegraphics[width=0.42\columnwidth]{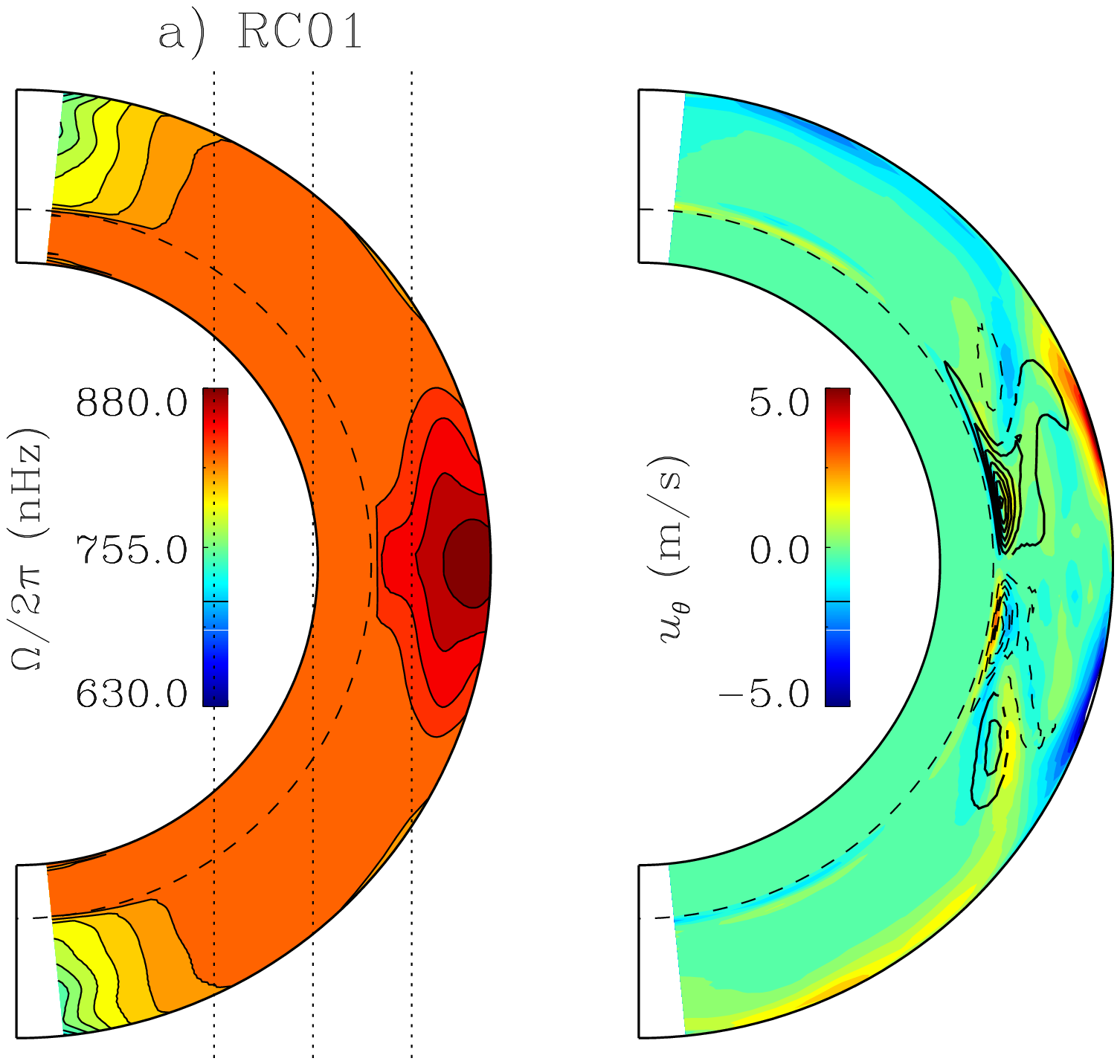}
\includegraphics[width=0.57\columnwidth]{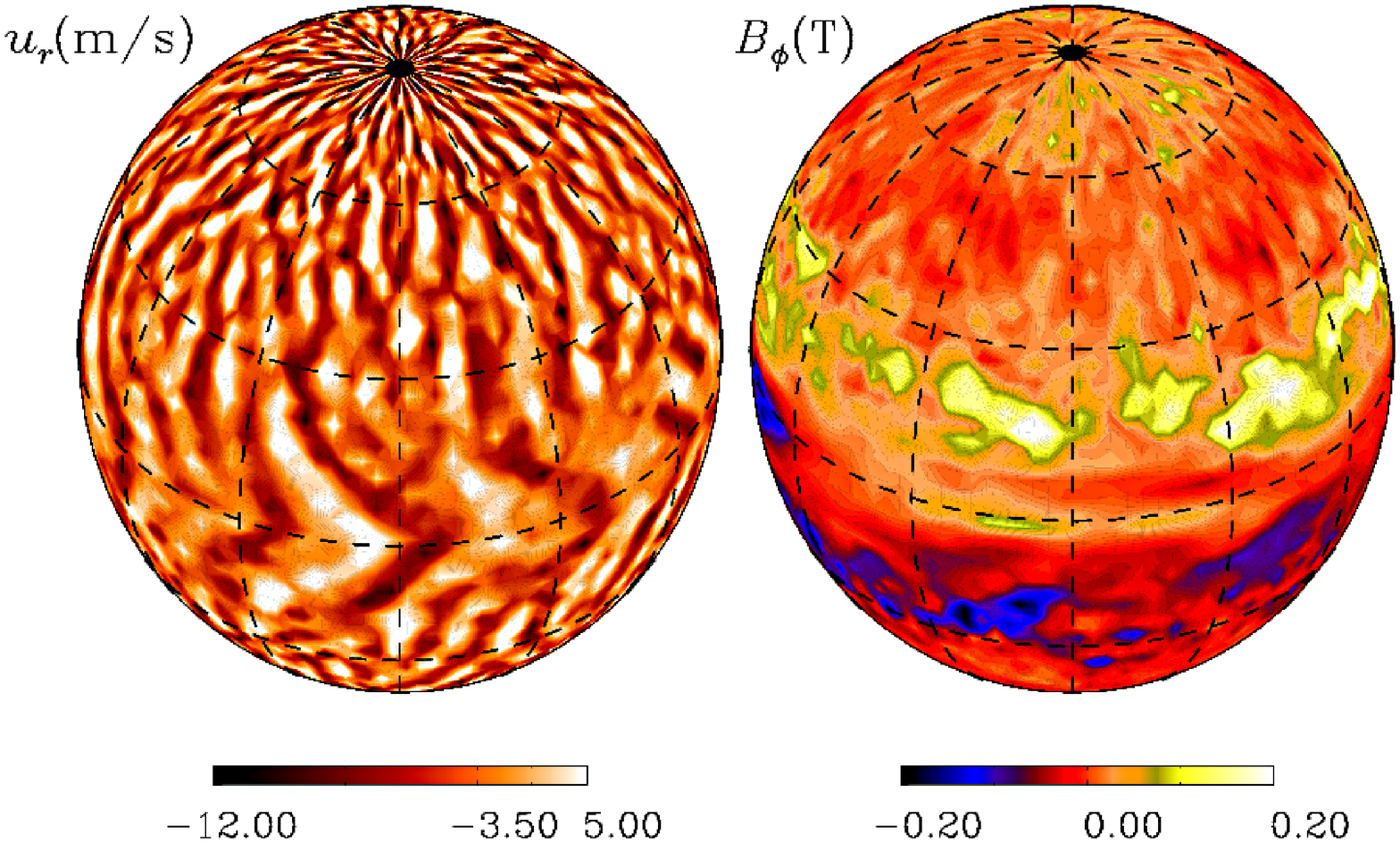}\\
\includegraphics[width=0.42\columnwidth]{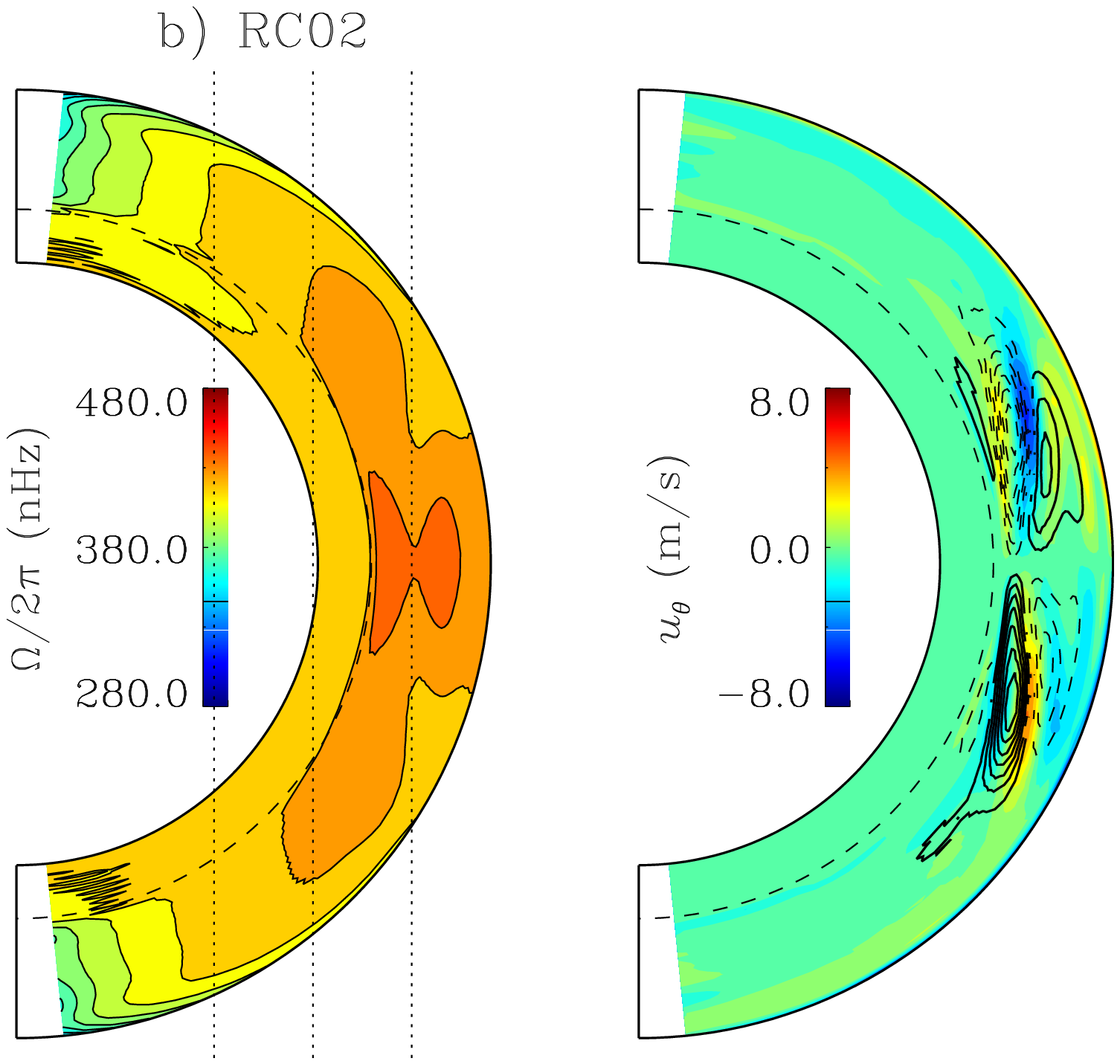}
\includegraphics[width=0.57\columnwidth]{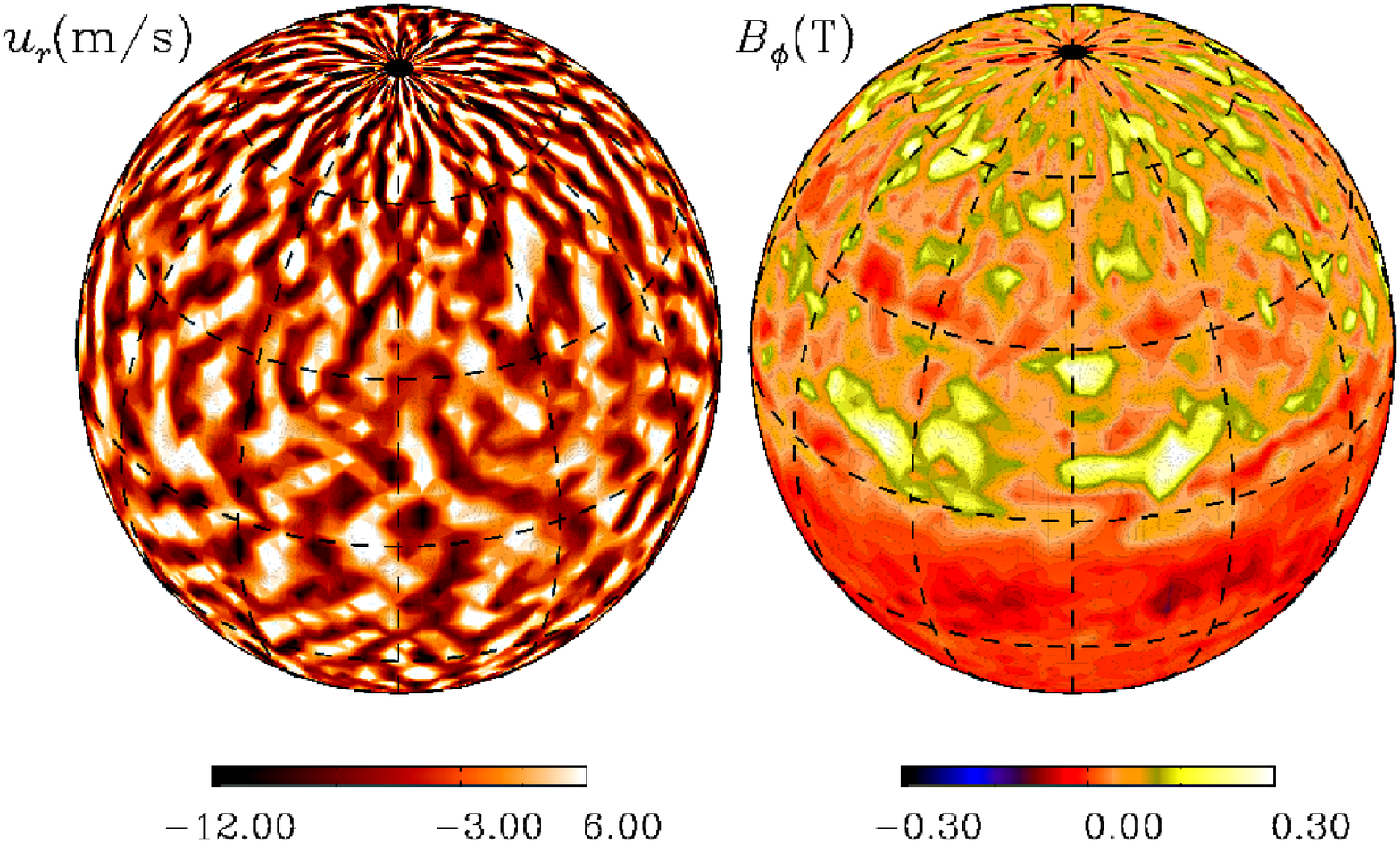}\\
\includegraphics[width=0.42\columnwidth]{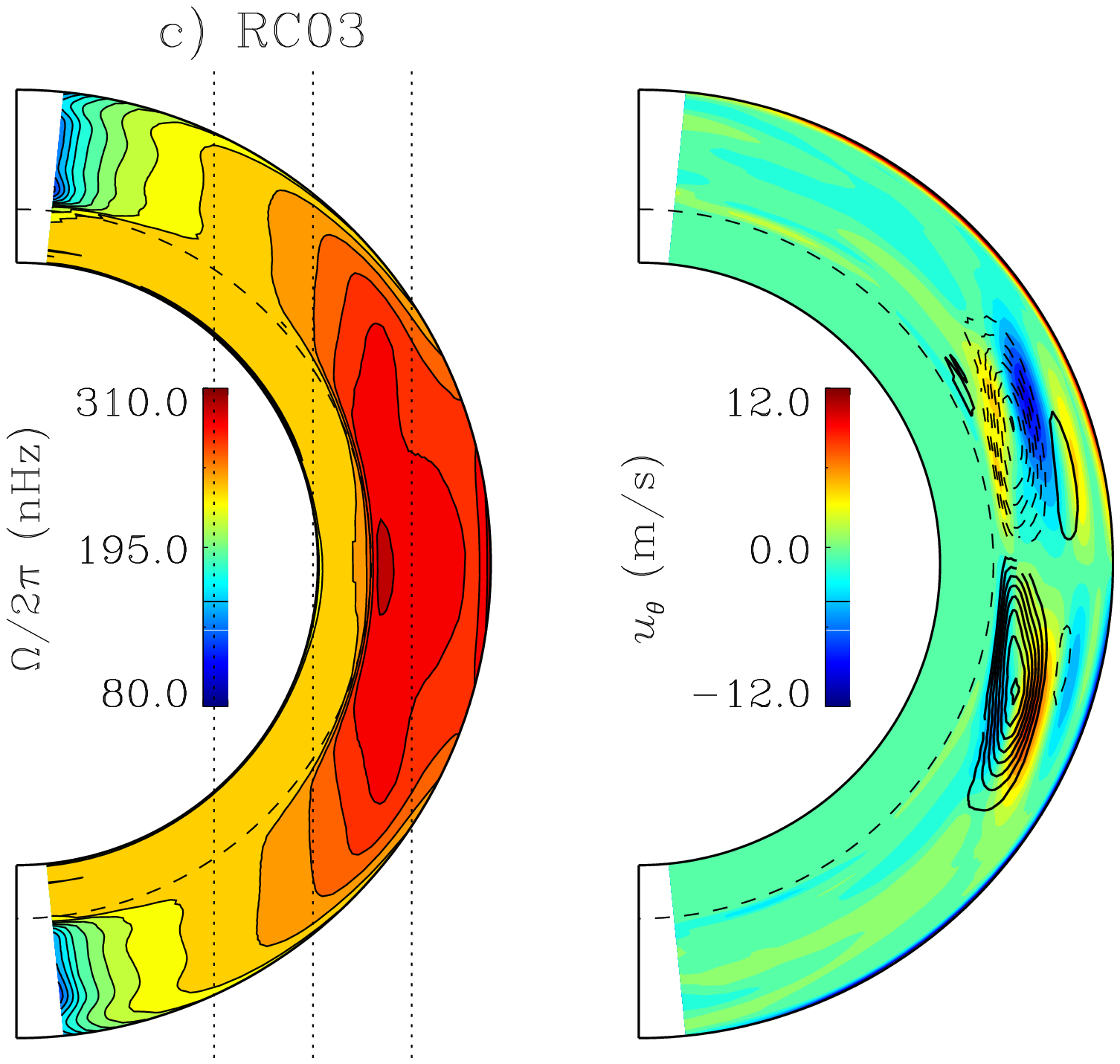}
\includegraphics[width=0.57\columnwidth]{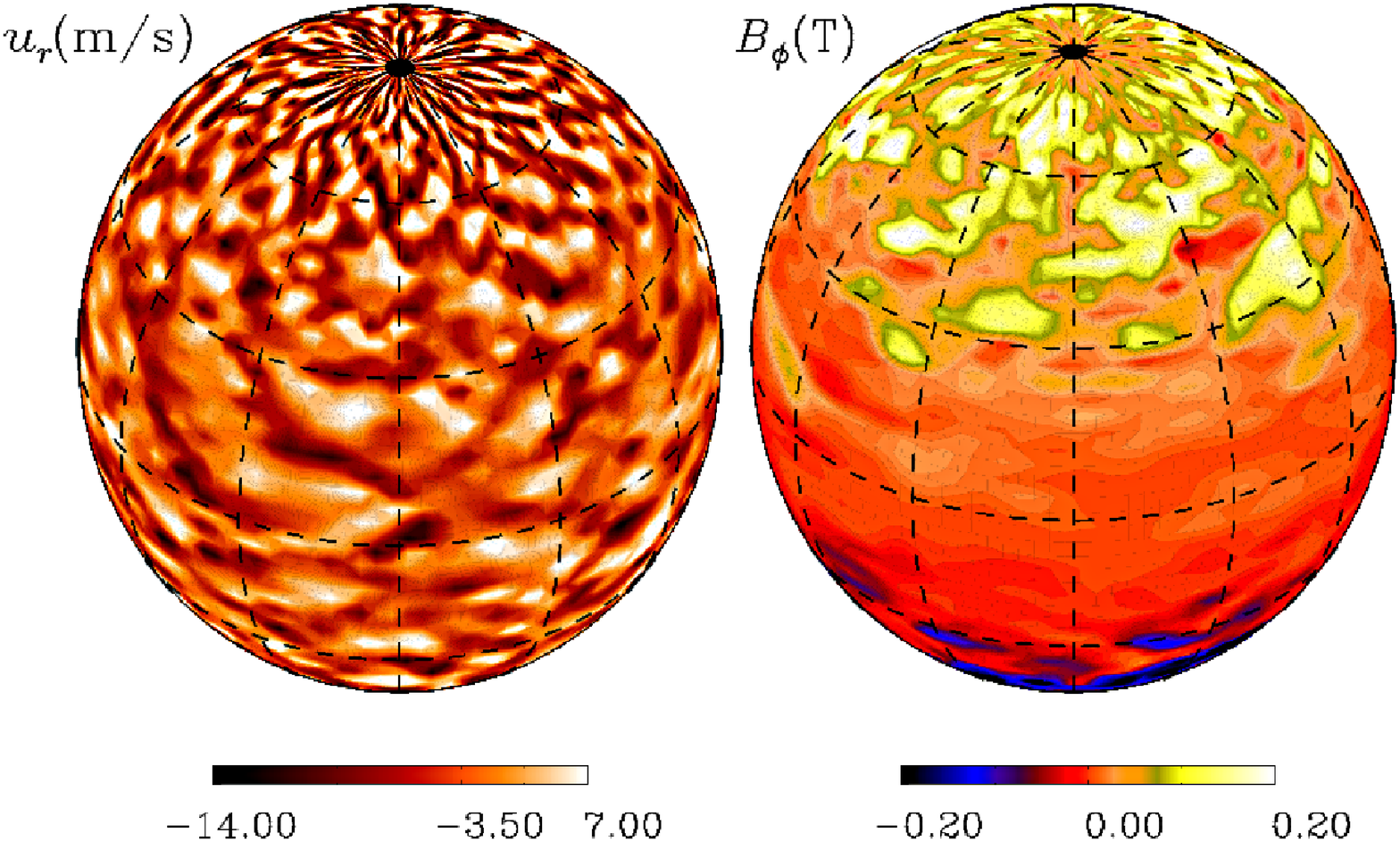}\\
\caption{Same as Fig. \ref{fig.df1} for the models a) ~RC01, b) ~RC02, 
and c) ~RC03. For these cases the temporal average is over $\sim 10$ years
during the steady state phase of the simulation. }
\label{fig.df2}
\end{center}
\end{figure}

It is worth mentioning first that due to the presence of the tachocline,
most of the magnetic field develops at the base of the convection zone.
Thus, the Maxwell stresses play an important role in the downwards 
transport of the angular momentum. In Figure \ref{fig.df2}, panels a, b and c show 
the differential rotation (left column)
and meridional circulation (right column) profiles of models ~RC01, ~RC02 and RC03,
respectively. Model ~RC01 has the lowest value of $\Ro$, thus
it has the strongest influence of the Coriolis force. 
This force tends to homogenize
the rotation of the convection zone with that of the radiative zone. 
This happens in a region between $\sim 5^{\circ}$ and $\sim 80^{\circ}$
latitude. However, zones with some radial and
latitudinal shears are developed at the equatorial and polar latitudes. 
In model ~RC02, both radial and latitudinal
differential rotations develop in larger zones. The intermediate latitudes 
are in iso-rotation with the radiative zone.
 
The transport of angular momentum to the radiative zone makes
the stable layer to rotate, on average (cf. Fig. \ref{fig.df2}b), faster than 
the frame. Hence, the equatorial acceleration of the convection zone in the
model RC02 is not as pronounced as in the model CZ02. In the former case
the angular velocity is around 440nHz, while in the latter case it reaches
500nHz. Consequently, the latitude of iso-rotation with the frame is moved 
up to $\sim 70^o$.  In the model  CZ02 it is located at $\sim 50^o$ (the 
yellow filled contours in Figs. \ref{fig.df1}b and \ref{fig.df2}b correspond 
approximately to the frame rotation rate, $\Omega_0$).

Besides affecting the average profile of the differential rotation
(with respect to purely hydrodynamic models)
the magnetic feedback on the flow generates torsional oscillations in 
the models where the dynamo is periodic (CZ02 and RC02). 
The latitudinal morphology of the simulated oscillations resemble the 
solar observations, but the amplitude of 
the observed oscillations is a few times smaller. This might 
be a consequence of a rather large $\alpha$-effect such as found by
\cite{CMT04}.  The origin of the torsional oscillations in our 
simulations seems to be due to a modulation of the latitudinal 
angular momentum transport mediated by the meridional circulation 
and the magnetic torque (at equatorial latitudes). 
A comprehensive discussion of these results is beyond the scope 
of the current work and is presented in a separate paper \citep{GSDKM16b}. 

A NSSL, more pronounced at higher latitudes, is also observed 
in this model (RC01). 
In both ~RC01 and ~RC02 cases the contours of rotation form conical shapes 
in the convection zone (see the vertical dotted lines in 
Fig. \ref{fig.df2} as an eye guide). Finally, in model ~RC03 
there is a well defined latitudinal differential rotation 
in the bulk of the convection zone, and also a NSSL is formed. The region of 
iso-rotation of the convection zone with the radiative core occurs at higher 
latitudes ($\sim 70^{\circ}$ latitude). For this reason the radial shear
at the equator, in the tacocline, is stronger in model ~RC03 than in models 
~RC01 and ~RC02.
The contours of model ~RC03 appear 
cylindrical which is surprising because in this model the Coriolis force
influence is smaller than in the other two cases. The rightmost 
columns of Fig. \ref{fig.df2} show the convective structure, represented
by the vertical velocity and the distribution of the
toroidal field at $r=0.95\Rs$.

\begin{figure}[hbt]
\begin{center}
\includegraphics[width=0.49\columnwidth]{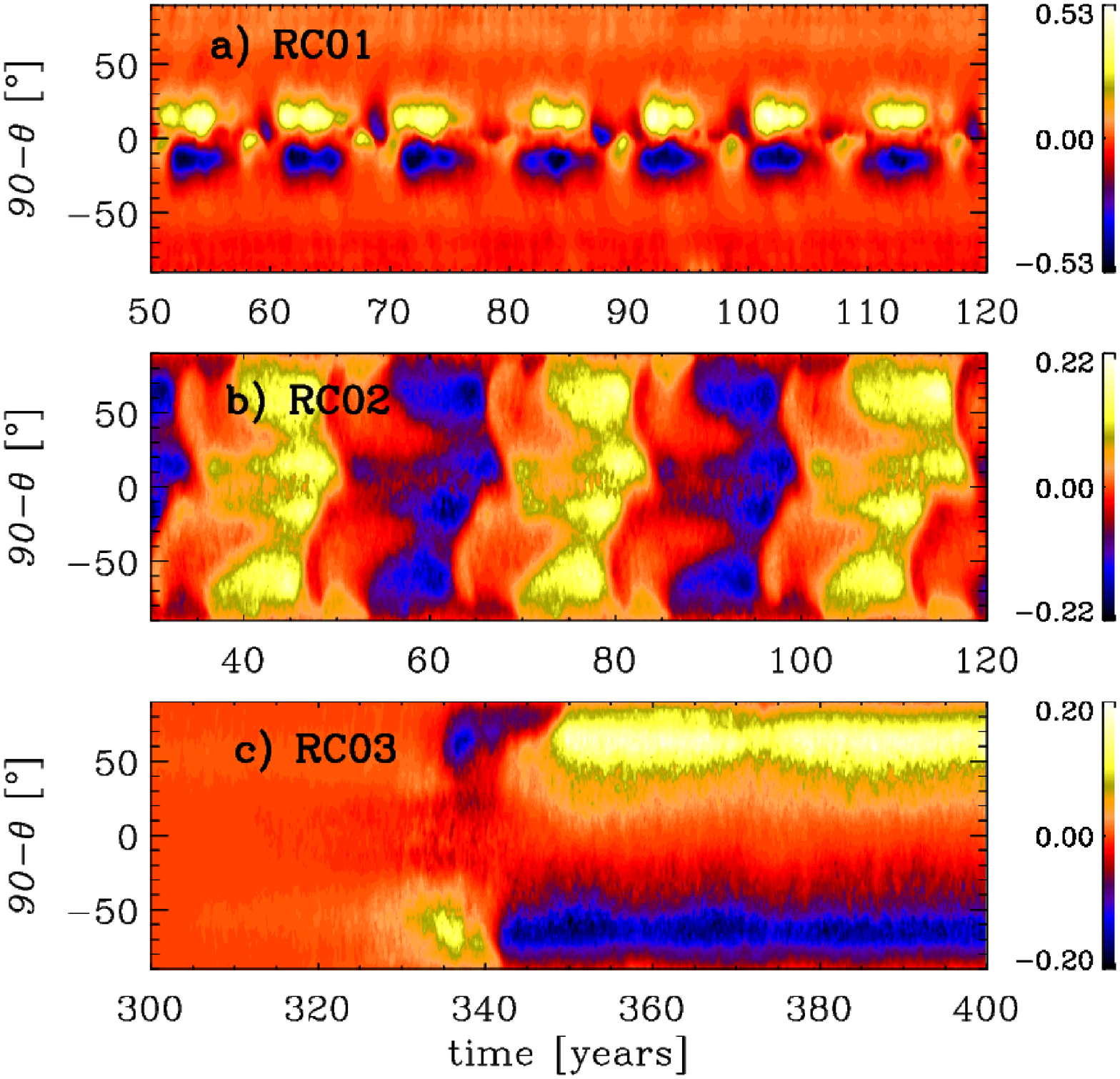}
\includegraphics[width=0.49\columnwidth]{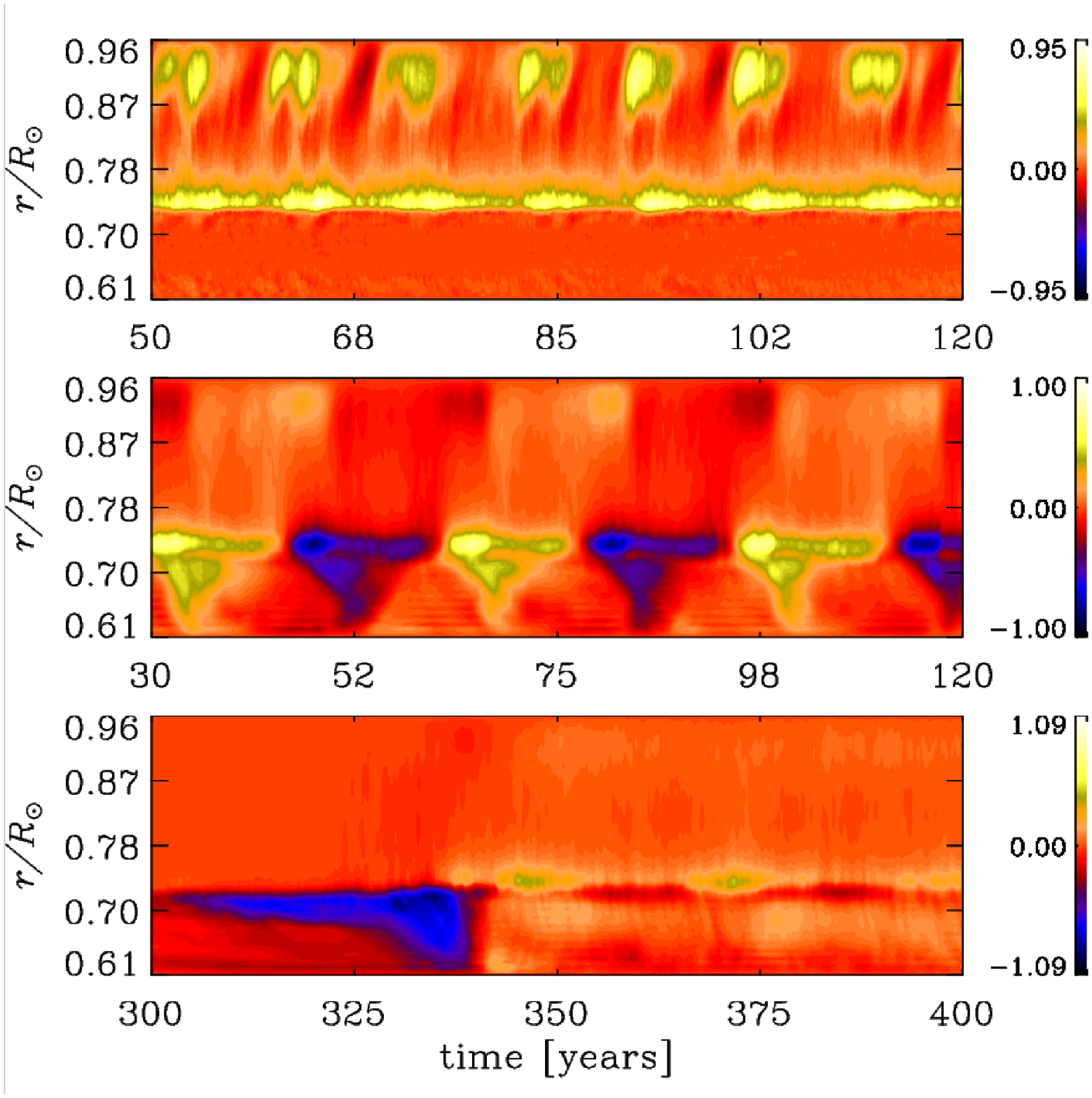}
\caption{Same than Fig. \ref{fig.bd1} for the models a) ~RC01, b) ~RC02 
and c) RC03. The color scales show $\mean{B}_{\phi}$ in Tesla.}
\label{fig.bd2}
\end{center}
\end{figure}

The meridional circulation is multicellular in all our models with the
tachocline. However, as expected, model ~RC03 has larger meridional 
velocities than the other models
and exhibits a dominant counterclockwise (clockwise) cell in the Northern
(Southern) hemispheres. Noteworthy, in all these models a low amplitude 
poleward flow develops in the upper part of the convection 
zone at latitudes $>30^{\circ}$.  This can be noticed in 
the second column of Fig. \ref{fig.df2} in 
the color filled contours of the latitudinal velocity, $\mean{u}_{\theta}$, as the 
blue and yellow regions in the Northern and Southern hemispheres, respectively. The
formation of this flow is likely due to the gyroscopic pumping mechanism
because of the negative gradient of angular velocity \citep{MH11}. 
 
The third column of Fig.~\ref{fig.df2} shows that the smaller convective 
structures in our simulations are formed at higher latitudes. In the equatorial 
band the structures are elongated, resembling the so-called banana cells.
Note that the vertical domain of our 
simulations reaches up to $0.96\Rs$, and small-scale structures are not resolved. 
We surmise that this is the reason why NSSL and latitudinal meridional 
circulation only appear above $\sim 30^{\circ}$ latitude. 
 Nevertheless, the meridional flow structure qualitatively agrees with 
recent helioseismology inversions made using the Solar Dynamics 
Observatory data \citep{zbkdh13,str13}.

The magnetic field in these models evolves on longer timescales than
in the CZ models  (see \S\ref{sec.cp}). 
The fast rotating model, ~RC01, develops 
a magnetic field that oscillates in amplitude but does not show clear 
polarity reversals. The oscillation period is $\sim 10$ yr (not included 
in Table \ref{tbl.1} because it does not correspond 
to magnetic field reversals).  
The topology of the field consists of 
wreaths of toroidal field of opposite polarity across the equator. 

\begin{figure}[H]
\begin{center}
\includegraphics[width=0.69\columnwidth]{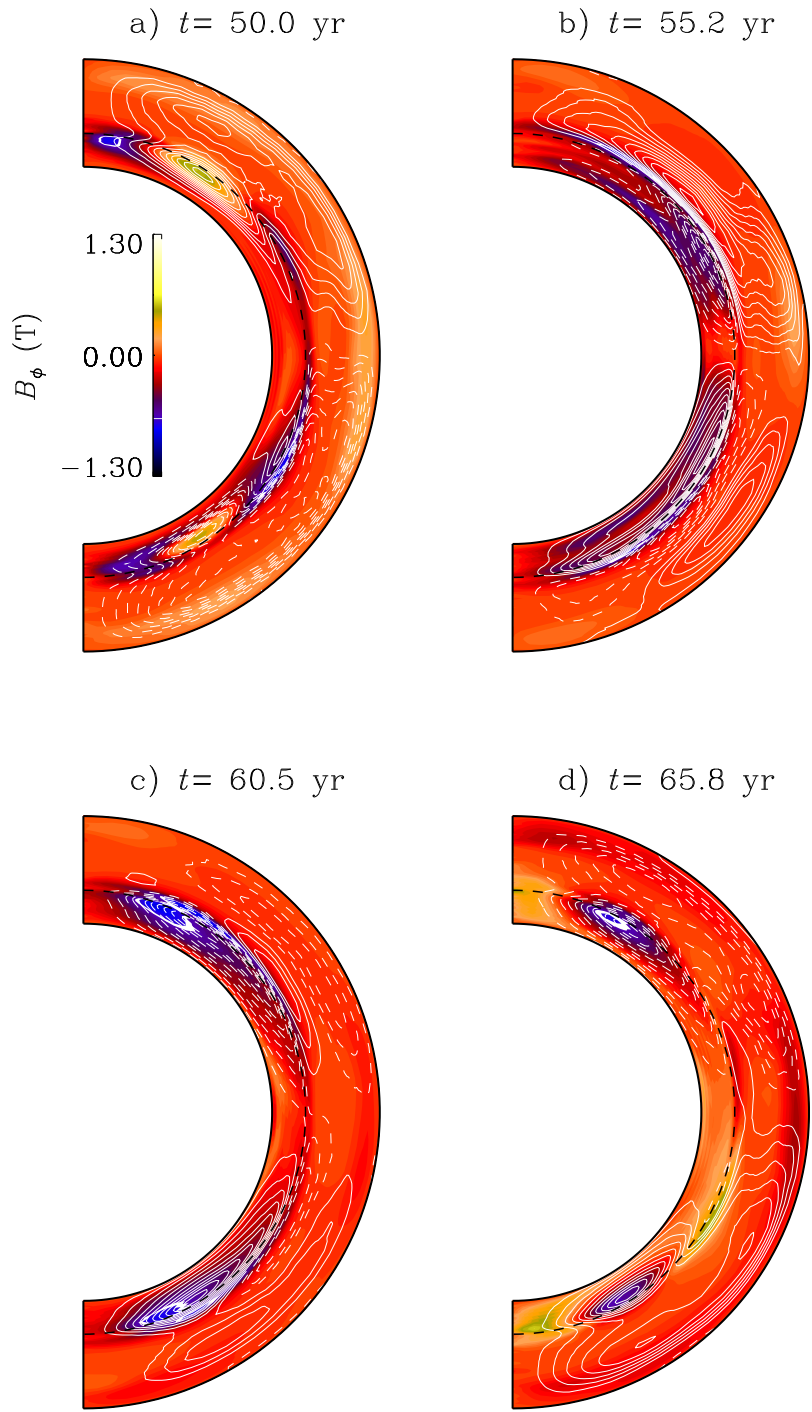}
\caption{Same than Fig. \ref{fig.revnt} for model RC02.}
\label{fig.revt}
\end{center}
\end{figure}

Model ~RC02 presents magnetic cycles with a full period of $\sim 30$yr.
Unlike the solar magnetic field, in this model the toroidal (poloidal) 
component of the magnetic field is symmetric 
(anti-symmetric) relative to the equator (see Figs. \ref{fig.bd2}b 
and \ref{fig.revt}). 
%The magnetic to kinetic energy ratio is one order of magnitude 
%larger than in the corresponding model without tachocline, ~CZ02. 
Most of the magnetic energy in this case is in the 
toroidal component of the field.  
In the simulation with the slowest rotation rate (model ~RC03), 
the magnetic field is nearly  steady (as in ~CZ03) with a larger 
concentration of $\mean{B}_{\phi}$ in a narrow region at the base 
of the convection zone at the equatorial latitudes. 
The field at the surface has the same polarity as at the bottom
and is concentrated at higher latitudes.

An important disparity between the models ~RC and their non-tachocline
counterparts, models ~CZ, is the ratio between their volume averaged
magnetic and kinetic energies as can be seen in Table \ref{tbl.1}. 
This difference is absent in the surface layers where both sets of 
models show magnetic fields of similar magnitude (in fact, models CZ have 
slightly higher values, see 
Table \ref{tbl.2}).  
In both, CZ and RC cases, the amplitude of the surface field correlates 
directly with the rotation rate. Nevertheless, because of the radial shear 
at the tachocline, in the ~RC models  the maximum mean-field in that region 
reaches up to $\sim 1$ T ($10^4$ Gauss), as evident in the right panels
of Fig. \ref{fig.bd2} and Table \ref{tbl.2} (values in parenthesis). 

As for the turbulent coefficients in these models, there is a clear
correlation between the turbulent diffusion coefficient, $\etato$, 
and the rotation rate, indicating that
rotation quenches turbulent diffusion. However, while $\Ro$ changes
by a factor of $\sim 5$ between the faster and slower models, $\etato$
changes only by a factor of $\sim1.15$.  Similar to the models CZ, 
there is no 
correlation between the kinetic $\alpha$ effect and the rotation.
Figure \ref{fig.alpha-omega}b shows that the radial 
profiles of $C^k_{\alpha}$ are roughly the same for the three models.
They have small positive values at the radiative zone, reach
a positive maximum in the middle of the convection zone and
change sign near the upper boundary.
On the other hand, $C_{\alpha}^m$ changes proportionally to the 
rotation rate. The sign of the $C_{\alpha}^m$ at the tachocline
is positive, and it is the dominant term in the total $\alpha$-effect 
in that region. In the lower/middle part of the convection zone the sign of
$C_{\alpha}^m$ is opposite to $C_{\alpha}^k$, but its amplitude 
is small. For models ~RC01 and ~RC02 in the upper convection 
zone  $C_{\alpha}^m$ is negative and, because of the low density, 
reaches maximum values. Similarly to models CZ, both 
$C_{\alpha}^k$ and $C_{\alpha}^m$ 
contribute to the field generation. For the slowest  
rotating  model, ~RC03, the contribution of $C_{\alpha}^m$ is
unimportant. 

Models RC develop strong radial shear at the interface between
the radiative and convective layers (Fig. \ref{fig.alpha-omega}d). This
radial shear anti-correlates with 
the rotation rate, i.e., the
slower the rotation rate, the largest the shear at the tachocline. 
Similar to the solar case, the radial shear at the tachocline is negative at
the poles and positive at the equator.  According to helioseismology the 
Sun exhibits the strongest shear at the poles, whereas all the models in this 
section have the strongest shear at the equator.  The profiles of $\partial_r \Omega$ 
are fairly flat in the convection zone and negative in the NSSL at
high latitudes (Fig. \ref{fig.alpha-omega}d). 
As for the latitudinal differential rotation, the fast rotating 
model shows a peak at low latitudes (between $15^{\circ}$ and $40^{\circ}$ 
latitude) followed for a flat, zero
shear, region in middle latitudes and larger gradients of 
$\Omega$ in latitudes above $60^{\circ}$ (see red line in Fig. 
\ref{fig.alpha-omega}f). Models ~RC02 and ~RC03 have 
profiles which are comparable at lower and middle latitudes.
However, near the poles, the profile of model ~RC03 shows a 
stronger latitudinal shear. 

The morphology of the steady magnetic field in models RC01 and 
RC03 can be described in terms of the distribution of its source terms.
For model RC01, at the tachocline, both sources of magnetic field
are small, the dynamo results in a steady field  
concentrated at the equator where $\partial_{\theta}\Omega$ peaks 
(red dashed line in Fig. \ref{fig.alpha-omega}f).
In this region the $\alpha$-effect is possibly dominant over
the $\Omega$-effect, yet the poloidal component of the field
is $\sim 4$ times larger than the toroidal one. 
In the upper convection zone, 
$\partial_{\theta}\Omega$ has larger values at the equator, which
explain why the field is concentrated in a band at 
$\pm 30^{\circ}$ latitude (Fig. \ref{fig.bd2}a). 
The field variations do not lead to 
polarity reversals, perhaps because 
the field at the tachocline remains steady. 
For model ~RC03 the field reaches its saturation after $t\simeq340$ yr
with steady magnetic fields.
Since in this case $\partial_{\theta}\Omega$ reaches larger values at 
the poles, the toroidal magnetic field is localized at polar latitudes. 

Unlike the oscillatory models  without the tachocline, ~CZ01 and ~CZ02, 
the magnetic field migration in model with the tachocline ~RC02 cannot be solely
explained in terms of the Parker-Yoshimura rule. Because of the 
long cycle-period of activity, other transport processes like the meridional 
circulation or turbulent pumping can affect the 
magnetic field evolution. At the tachocline the migration
of the field can be explained by the Parker-Yoshimura rule (Eq. \ref{eq.py}).
The field developed at
the poles migrates equatorward, and the field developed at the 
equator migrates poleward. This migrating process can be seen
in Fig. \ref{fig.revt}a-d which, similarly to Fig.~\ref{fig.revnt}, 
shows a half of the dynamo cycle with one magnetic field reversal 
illustrates in four snapshots. These dynamo
waves result from a positive $\alpha$ and negative (positive)
values of $\partial_{r} \Omega$ at higher (lower) latitudes. 
The upward radial migration from $r\simeq0.75\Rs$ to $r\simeq0.90\Rs$ 
is also in agreement with Eq. (\ref{eq.py}) for 
$\alpha \partial_{\theta} \Omega > 0$ in this region. In the
upper convective layer, above $r\simeq0.90\Rs$, there is a 
region of strong
magnetic field, possibly resulting from the large negative
radial shear. The latitudinal migration in this region, depicted
in Fig. \ref{fig.bd2}b, shows two branches of slighty equatorward
migration: one branch is located between $0^{\circ}$ and $\pm 30^{\circ}$ 
latitude, the other one is above  $50^{\circ}$ latitude. 
According to Eq. (\ref{eq.py}) these branches would 
require negative values of ${\bm s}$; 
however this does not agree with the profiles of $\alpha$ and 
$\partial_r \Omega$ shown in Fig. \ref{fig.alpha-omega}b and 
\ref{fig.alpha-omega}d (blue lines). 
The most equatorial branch of $\mean{B}_{\phi}$ coincides with a 
clockwise meridional circulation cell with an equatorward flow at the surface 
(Fig. \ref{fig.df2}b). However, it is impossible to assure that
this is the case giving the difficulty in disentangling 
the local and non-local effects contributing to the field migration. 

\subsection{What process governs the cycle-period?}
\label{sec.cp}
Perhaps the most interesting difference between the oscillatory dynamo 
models  CZ02 and RC02,  
is the timescale of the magnetic cycles (i.e., the
period of the magnetic polarity reversals). As follows from 
the mean-field dynamo theory, the 
magnetic field grows and decays according to the values of the inductive
and the diffusive terms in the induction equation \citep{BS05}. 
A long standing problem for mean-field dynamo modelers has
been to conciliate the theoretically expected value of the turbulent 
diffusivity, $\etat$, with the observed cycle period. In the solar case
a mixing-length 
theory (MLT) estimation of $\etat$ gives values $\sim 10^9$ m$^2$s$^{-1}$ 
which, in turn, results in dynamo cycles with periods of $2-8$ yr \citep{GDDP09} 
instead of 
the observed $22$ yr. Non-linearities in the dependence of the turbulent
diffusivity on the large-scale magnetic field, like the so-called 
$\eta$-quenching mechanism, have been explored but did not  
solve the problem \citep{RKK94,GDDP09,MJNM11}. 
The cycle-period difference between the two types of models found
in our simulations provides an opportunity to explore this issue.  

\begin{figure}[H]
\begin{center}
\includegraphics[width=0.69\columnwidth]{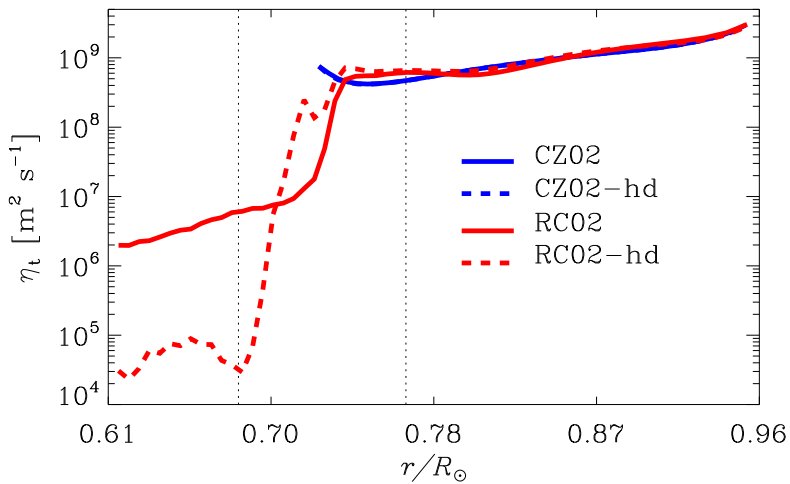}
\caption{Vertical profiles of the turbulent magnetic diffusivity,
$\etat$, for model ~CZ01 (blue) and ~RC02 (red). The dashed lines
show the profiles of $\etat$ for the hydrodynamic versions of
models ~CZ02 and ~RC02.}
\label{fig.etat}
\end{center}
\end{figure}

The radial profile of the turbulent diffusion coefficient, 
$\etat$, computed using Eq.  (\ref{eq.etat}), for both 
models with the solar rotation rate, ~CZ02  and ~RC02,
is depicted in Fig. \ref{fig.etat}.  From this figure and the 
values of $\etato$ in Table \ref{tbl.2}, it can be seen that the 
profiles and values of $\etat$ in the convection zone 
match closely in both models. Both cases correspond to a 
strongly diffusive 
regime with $\etat\simeq10^9$ m$^2$s$^{-1}$ (in agreement with the
MLT estimate). However, model ~CZ02 has a $\sim 2$-yr cycle period, 
while model ~RC02 has a full magnetic cycle-period of $\sim 30$ years.
Furthermore, in the same figure, the dashed lines correspond to 
the turbulent diffusivity of the hydrodynamic versions of the CZ02 
and RC02 models. We notice that there is no important diffusivity 
quenching due to the presence of the saturated magnetic field 
except in a small fraction of the radius nearby the tachocline
where the diffusivity of the magnetic model is smaller (see the 
region between the vertical dotted lines in Fig. \ref{fig.etat}).

This suggests that the cycle period in model ~RC02 may be determined by 
the value of $\etat$ in the region where most of the magnetic 
field is produced and stored (i.e., at and below the tachocline), 
which is mainly convectively stable. (Note, 
that $\etat$ does not tend to zero in the radiative layer but to values 
around  $\sim10^6$ m$^2$s$^{-1}$. This reveals that the
magnetic field is inducing turbulent motions in some fraction of the 
radiative interior.) To verify if this is indeed the case, we have 
computed the ratio between the diffusion times for models RC02 and CZ02, 
as follows:
\begin{equation}
\frac{t_d^{RC}}{t_d^{CZ}}\simeq \frac{L^2_{RC} \brac{\etat}_{CZ}}{L^2_{CZ} 
\brac{\etat}_{RC}}\simeq 19 ~,
\end{equation}
where, $L_{CZ}=(0.96-0.72)\Rs$ is the length 
scale of the convective layer of model ~CZ02, and $L_{RC}=(0.77-0.68)\Rs$ is 
the length of the  region of model ~RC02 where most of magnetic field 
is stored (see vertical dotted 
lines in Figs \ref{fig.alpha-omega}a and b); $\brac{\etat}_{CZ}$ and
$\brac{\etat}_{RC}$ are the radially averaged values of the turbulent diffusivity
over the same lengths.  Interestingly, this rough estimation agrees fairly
well with the ratio between the cycle periods of these two types of 
models.

It is striking that in model ~RC02 the magnetic 
field in the turbulent convection zone, despite the larger local values of 
$\etat$, evolves on the same timescale as the magnetic field in the most stable 
layer with lower diffusivity. If such a  non-local mechanism is operating in 
the Sun, it could explain the 22-year
period of the solar magnetic cycle.  This does not necessarily mean, 
however, that the migration 
of the sunspots observed at the surface is shaped by the meridional
circulation at 
the base of the convection zone as assumed by most of the flux-transport dynamo models.
On the contrary, the dynamo in our simulations operates distributed over 
the entire convection zone, and several processes could 
influence the magnetic field generation and migration 
(e.g., the near-surface shear
rotational shear layer, as argued by \cite{B05,PK11}). 

\subsection{Tachocline instabilities}

From Fig.~\ref{fig.etat} it is evident that turbulent motions must be present
in the stably stratified layer. Such motions increase the turbulent diffusivity 
by more than one order of magnitude compared to the pure hydrodynamic  
model RC02 (red dashed line).  The most probable origin for this hydromagnetic 
turbulence in the stable layer is the development of MHD 
instabilities at and below the interface
between the convective and the radiative layers.  It has been suggested that these
turbulent motions could be significantly helical such that, similarly to the
flow in the convection zone, they can result in a kinetic $\alpha$-effect 
\cite[e.g.,][]{DG01}. The analysis performed in \S\ref{s.tac}
indicates that the kinetic part of the $\alpha$-effect in the stable layer
is small when compared  with its values at the convection zone. Nevertheless, 
the magnetic contribution to $\alpha$ in the radiative zone is important, as
evidenced in Fig.~\ref{fig.alpha-omega}b. The development of small-scale 
helical magnetic structures due to instabilities could explain the existence of this 
magnetic $\alpha$-effect. This, in turn, leads to the generation
of large-scale magnetic field \citep{BS05}. 

This section does not intend to 
explore in detail the development of all possible tachocline instabilities in
the simulations presented here. Such analysis has been recently performed already 
by \cite{LSC15}
for one of the  EULAG-MHD simulations \citep{PC14}.  In this section we briefly 
review the nature of such instabilities, try to identify which of them could be operating 
in our simulations based in a simple energy analysis,  and discuss their 
contribution to the dynamo mechanism.  

Tachocline instabilities have been extensively investigated over the past decades. 
According to the energy source they can be divided into four distinct kinds which include
\citep{Arlt09}: a) shear-driven instabilities; b) baroclinic instabilities; 
c) buoyancy driven instabilities; and d) current-driven instabilities.
The shear-driven instabilities \citep[e.g.,][]{DK87,CDG99,DG01} are purely 
hydrodynamic. Due to the latitudinal shear at the tachocline, small perturbations 
can destabilize the flow given sufficiently strong shear, e.g., more than
$\sim30$\% difference between the equator and the pole for a 2D analysis. 
The 3D stability analysis 
of \cite{ASR05} has shown that the solar tachocline should be nearly stable, in
agreement with previous results by \cite{Garaud01}. In the simulations presented 
in \S\ref{s.tac}
the difference in the angular velocity between equator and pole is less than 10\%
for the models RC01 and RC02, and about 30\% for the model RC03.  It seems unlikely 
that this kind of instability is developing in these simulations. 
Baroclinic instabilities (b) are more complex in nature. If the transition region 
from subadiabatic to
super-adiabatic is in thermal wind balance, horizontal (azimuthal and latitudinal) 
perturbation flows can gain some thermal energy and become unstable. Recently
\cite{Gi15} has demonstrated that the magnetic field has a stabilizing effect
on these growing modes. Buoyancy driven instabilities (c) are presumably
responsible for the emergence of magnetic flux tubes. The 
buoyancy force of low density parcels of magnetized gas is  
responsible for the emergence. For toroidal field ropes stored in a
stable layer, it is generally true that the magnetic field should surpass
the equipartition field strength \citep[see][and references therein]{Hughes07}.  
This is not the case in the simulations above. Strong flux concentration
could be achieved in the case of very thin flux tubes which would require
higher grid resolutions. The current-driven instabilities (d) appear 
whenever there is a large-scale current
and their energy source is the large scale magnetic field.
This kind of instabilities could develop without rotation \citep[e.g.,][]{Tayler73,BU12} or
with rotation \citep{PT85}, or also when the rotation is  differential \citep[e.g.,][]{ASR05}.
These latter processes, often called magneto-shear instabilities \citep{CDG03},
destabilize different configurations of toroidal magnetic bands either
by opening the magnetic field lines (the so-called clamshell instability) or
by tipping  the axis of the magnetic field band. In both cases it is expected
the development of non-axi-symmetric magnetic field out of the axi-symmetric
field lines. This instability grows for different configurations of
toroidal magnetic fields and diminishes the shear. When, a latitudinal
shear profile and a toroidal field configuration are left to evolve
freely, after the saturation the instability decays \citep{CDG03}. However,
if the system is forced, as the solar tachocline should be, both the
shear and the magnetic field adjust to a new equilibrium state 
\cite[e.g.,][]{Miesch07}.

All the ingredients for the onset of the magneto-shear instability are 
present in the simulations with radiative zone presented above (RC models).
The differential rotation is sustained by the Reynolds stresses and,
depending on the Rossby number, it induces different configurations of 
toroidal fields (which in turn react back re-adjusting the shear). 
At some evolutionary stage of the simulation the conditions for the 
development of the instability are present creating small scale flows
and magnetic field in the stable layer. As a consequence, a new adjustment
should happen until the final steady state of the simulation is achieved. 

\begin{figure}[H]
\begin{center}
\includegraphics[width=0.8\columnwidth]{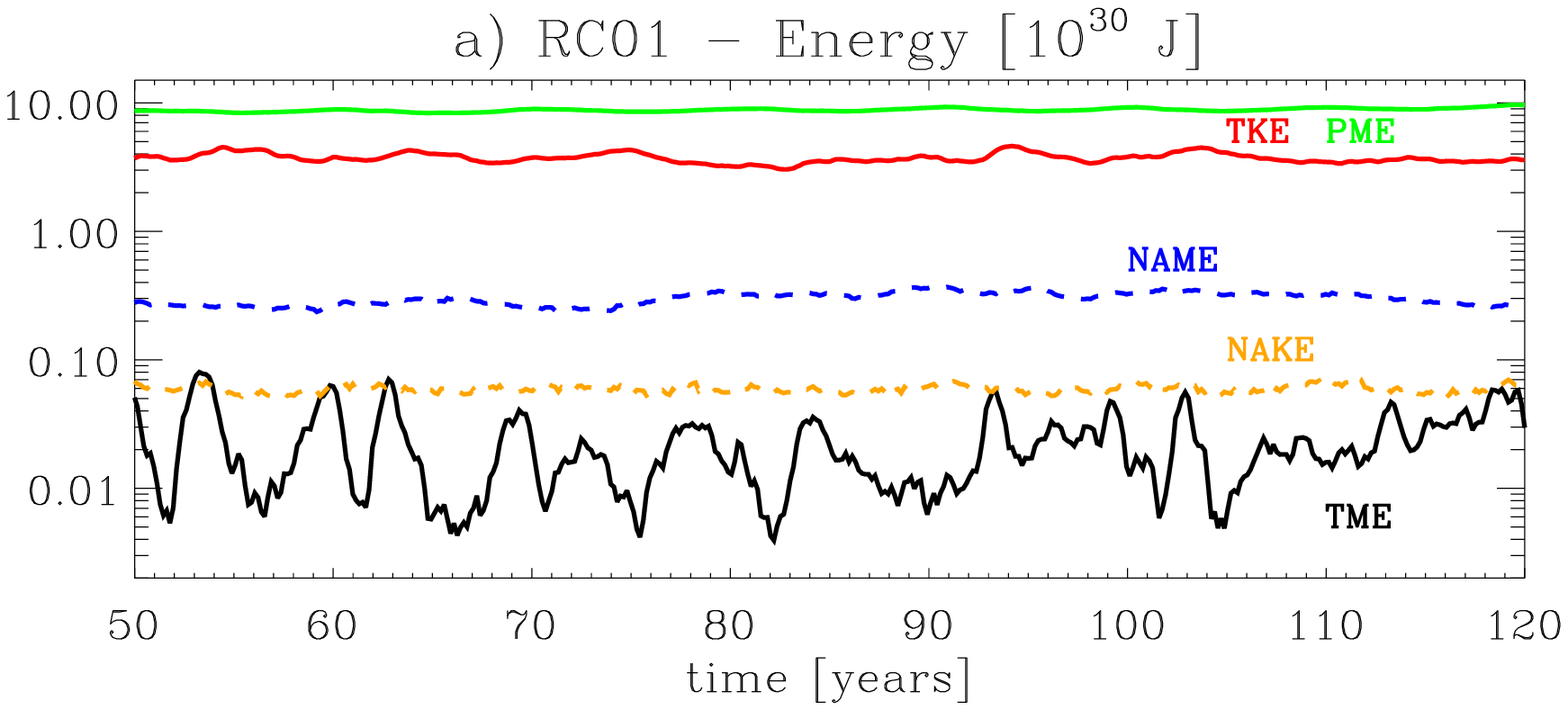}\\
\includegraphics[width=0.8\columnwidth]{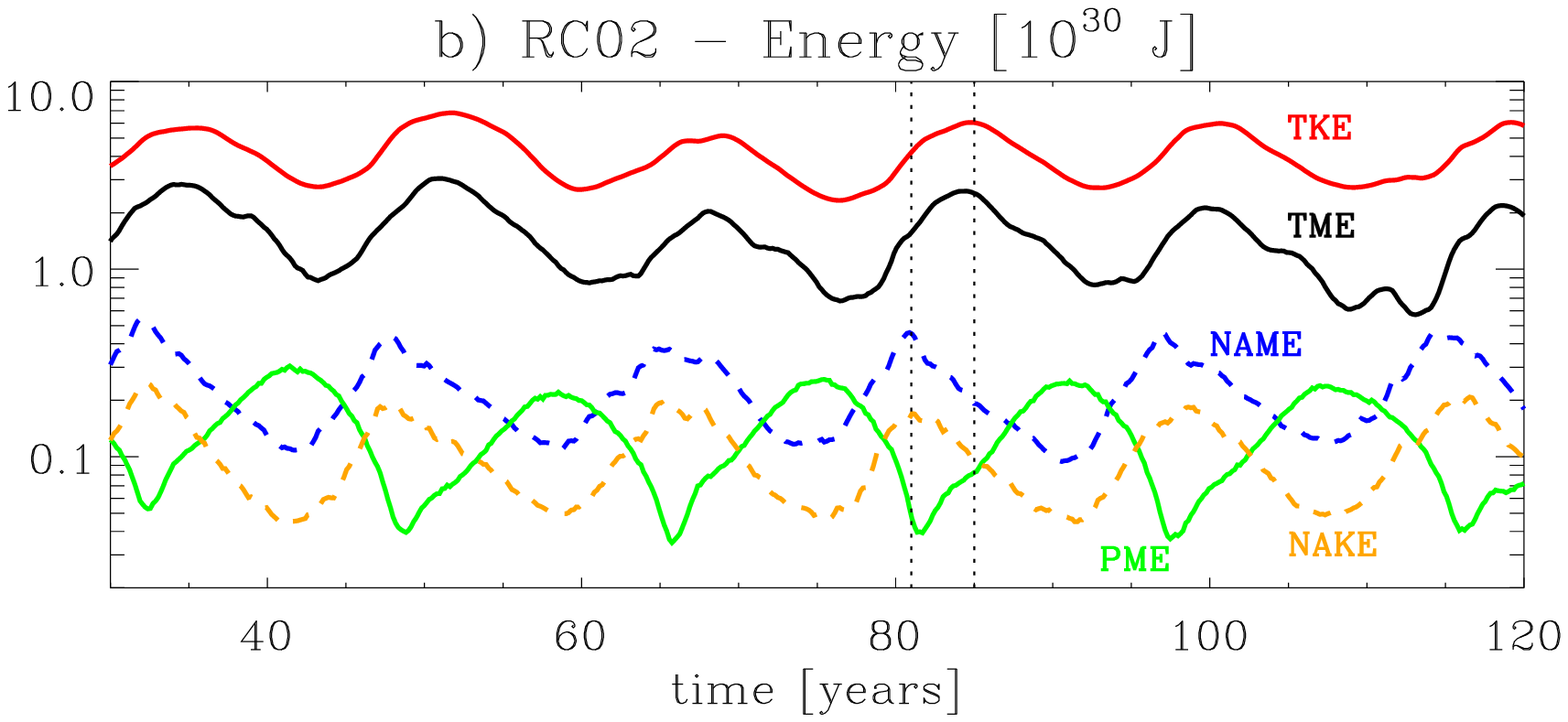}\\
\includegraphics[width=0.8\columnwidth]{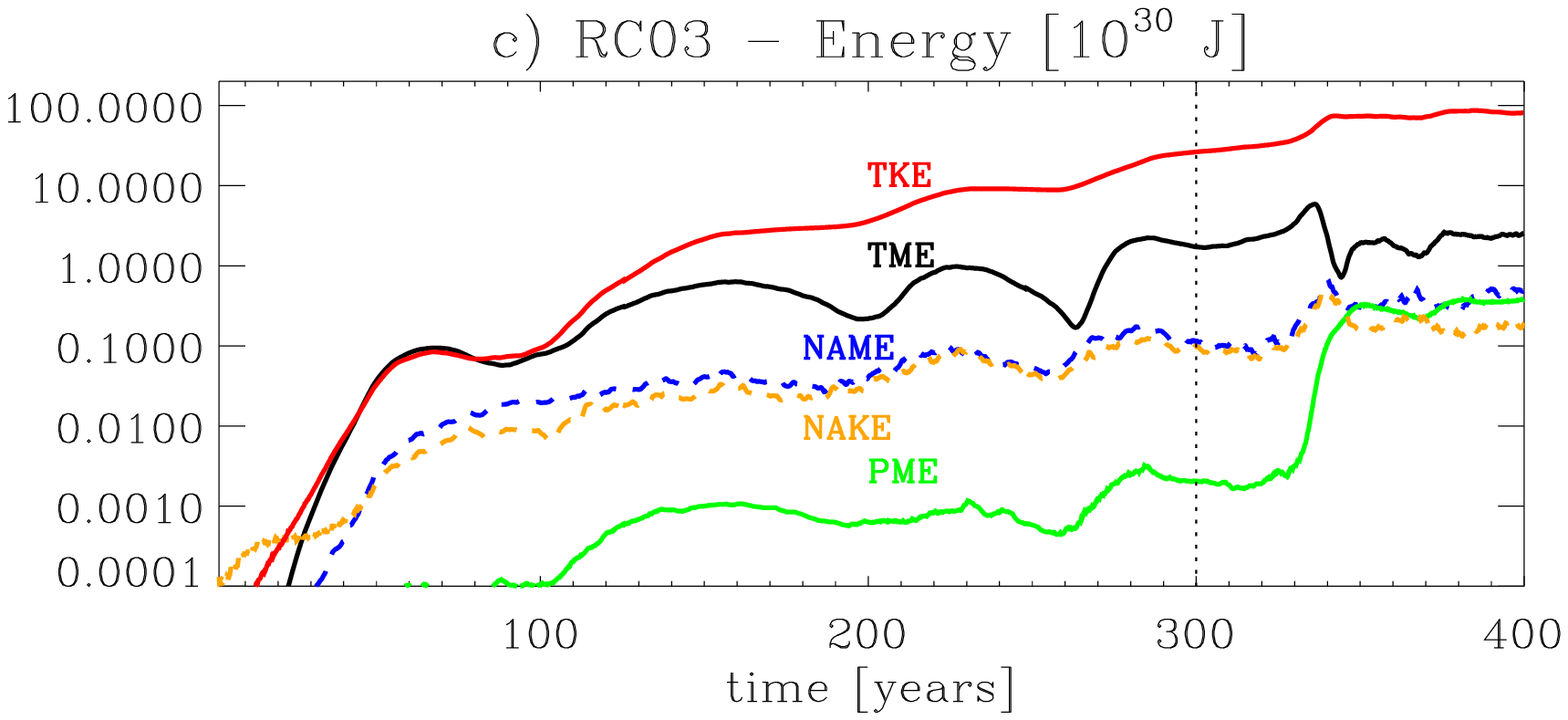}
\caption{Volume integral of the different energies in the stable region for
the models a) RC01, b) RC02 and c) RC03).
The toroidal magnetic energy (TME)  is shown in black, 
the toroidal kinetic energy (TKE) in red, the poloidal field energy (PFE) 
in green, the non-axisymmetric magnetic field (NAMF) with the dashed 
blue line and the non-axysymmetric kinetic energy (NAKE) with the 
dashed orange line. The dotted lines in the middle panels highlight the
phase-lag between the axi-symmetric and the non-axisymmetric quantities.
In the bottom panel, the dotted line shows the point from which the
butterfly diagram of Fig. \ref{fig.bd2}c was plotted.}
\label{fig.tinst}
\end{center}
\end{figure}

Following \cite{Miesch07}, we study the time evolution of the different energy
sources of the stable layer to verify the existence, and relevance,  of such small 
scale, non-axisymmetric, magnetic field.  The net energies are computed as follows 
\citep{LSC15}:

\begin{eqnarray}
{\rm TME} &=&  \frac{1}{2\mu_0} \int_{0.61}^{0.7} \int_{0}^{\pi/2} \int_0^{2\pi} 
	\mean{B}_{\phi}^2  r^2 \sin\theta d\phi d\theta dr ~, \\\nonumber
{\rm PME} &=&  \frac{1}{2\mu_0} \int_{0.61}^{0.7} \int_{0}^{\pi/2} \int_0^{2\pi} 
             (\mean{B}_r^2 + \mean{B}_{\theta}^2) r^2 \sin\theta d\phi d\theta dr ~,  \\\nonumber
{\rm NAME} &=&  \frac{1}{2\mu_0} \int_{0.61}^{0.7} \int_{0}^{\pi/2} \int_0^{2\pi} 
	{\bm b}'^2  r^2 \sin\theta d\phi d\theta dr ~, \\\nonumber
{\rm TKE} &=&  \frac{1}{2} \int_{0.61}^{0.7} \int_{0}^{\pi/2} \int_0^{2\pi} 
        \rho \mean{u}_{\phi}^2  r^2 \sin\theta d\phi d\theta dr ~, \\\nonumber
{\rm NAKE} &=&  \frac{1}{2} \int_{0.61}^{0.7} \int_{0}^{\pi/2} \int_0^{2\pi} 
	\rho {\bm u}'^2  r^2 \sin\theta d\phi d\theta dr,
\end{eqnarray}

\noindent
where TME, PME and TKE, are the axi-symmetric toroidal and poloidal magnetic
energies and the azimuthal kinetic energy, respectively.
NAME and NAKE stand for non-axisymmetric magnetic and kinetic energies. 
The resulting evolution profiles
are presented in Fig. \ref{fig.tinst}.  The  axi-symmetric 
toroidal (poloidal) 
magnetic and kinetic energies are depicted by the continuous black (green) and 
red lines, the non-axisymmetric magnetic and kinetic energies are shown
by the dashed blue and orange lines.  The results indicate that all models 
develop non-axisymmetric motions and magnetic field in the stable layer, however
the energy distribution is different from model to model. Models RC01 and 
RC02 reach steady state after few years of evolution. In the first case the
toroidal magnetic field is weak and erratic in the radiative layer
(see right panel of Figs. \ref{fig.bd2}a and \ref{fig.tinst}a).  The weakness
of this component of the field
in the stable layer probably results from the inhibition of
penetrative convection by the fast rotation.  The poloidal field has larger
energy, which is consistent with the presence of a non-zero $\alpha$ effect
in this region (red line in Fig. \ref{fig.alpha-omega}b).  

The case of model RC02 is more interesting since there is an important
mean oscillatory toroidal field. The non-axisymmetric magnetic 
field is also oscillatory and exhibits a phase lag with respect to the
toroidal field (see the vertical dashed lines indicating particular peak
times for TME and NAME). This behaviour is similar to the one obtained by 
\cite{Miesch07,LSC15} who observe the growth and 
decay of these quantities. Nonetheless, in these works the values of 
NAME and TME are similar to each other and the phase lag is about $\pi/2$.  
In the model RC02 the 
toroidal energy is larger than the turbulent magnetic energy. 
It is noteworthy the fact that the mean poloidal field oscillates
with a $\pi/2$ phase lag with respect to the non-axisymmetric magnetic
field (i.e., when the PME grows the NAME decays).  Note also that the PME 
is smaller than the NAME which is suggestive of a magnetic $\alpha$-effect 
giving rise to a large-scale poloidal magnetic field. The energy of the
turbulent motions oscillates with the same phase than the turbulent
field but with lower amplitude.  Finally, note also that the mean kinetic
energy is larger than the toroidal energy, the figure indicates acceleration 
and deceleration of the angular velocity in phase with the toroidal 
magnetic field \citep{GSDKM16b}.

The model RC03 is the one that takes longer to reach steady state.
Since this case is the one that exhibits more variation, 
in Fig. \ref{fig.tinst}c we present the entire evolution of the energies
in the stable layer. The figure
shows that the small scale non-axisymmetric kinetic and magnetic field 
energy components develop after the
development of the toroidal field and that the evolution of these
two quantities is closely tight (see dashed lines). The mean toroidal 
kinetic energy evolves together with the mean toroidal field. When the
non-axisymmetric modes develop, the toroidal kinetic energy grows.
We interpret this behaviour as the inwards transport of angular 
momentum, which makes the stable region to accelerate. The poloidal
magnetic energy also develops after the establishment of MHD
turbulence in the radiative zone.  After around 350 years of evolution
the rotation seems to achieve a steady state. This modifies all
the other quantities, specially the mean poloidal energy, which
grows about 2 orders of magnitude in few years. Note that the
$\alpha$-effect has important positive values in the entire 
stable layer (see the profile of $\alpha$ for the model RC03, 
green line in Fig.~\ref{fig.alpha-omega}b which was computed 
in the final phase of the evolution).   

In general, it is evident that the dynamics of the convectively 
stable stratified layer increases in complexity for models with
larger Rossby number.  This, of course, means a dependence on the
unstable region where phenomena like penetrative convection or
turbulent pumping are established. The non-axisymmetric flow and 
field seem to be contributing to the generation of poloidal field.
From the FOSA analyses presented above, it is the
small scale magnetic field which is contributing through the magnetic 
$\alpha$-effect.  The turbulent flows in the radiative zone 
increase the turbulent diffusion in this region. For the oscillatory
model RC02, it defines the cycle period of the large-scale magnetic 
field in the entire domain.

\section{Summary and conclusions}
\label{s.con}
We have performed global 3D MHD simulations of turbulent convection
in rotating spherical shells using the anelastic approximation. 
We have used the code EULAG-MHD which captures the
contribution of the non-resolved scales via an implicit sub-grid
scale model. The main goal of this work was the comparison of the properties
of global dynamos for models of the convective 
envelope only (CZ models),  and models that include a part of the 
radiative core below the convection 
zone (RC models).  The stratification is set in such a way that 
the Rossby number ($\Ro$) is roughly the same in corresponding CZ and RC models.
These 
simulations aimed at contributing to a broader understanding of the 
amplification and dynamics of large-scale magnetic fields 
in the Sun and solar-type stars.  
Our main conclusions are summarized and briefly discussed below.

The resulting differential rotation profile depends on the balance between
the Coriolis, 
buoyancy and Lorentz forces. All the cases presented here develop a solar-like
differential rotation (with faster rotation at the equator and slower rotation
at the poles). 
This result differs from the purely
hydrodynamic simulations for which the anti-solar rotation type is 
found as a robust feature
of slow rotating convection 
\citep{Gi76,GG82,SF07,MDBB11,KMGBC11,GSKM13b,GYMRW14}. 

Nevertheless, our results are in agreement with recent global models that 
obtain the transition from anti-solar to solar-like rotation due to the presence 
of dynamo-generated magnetic fields \citep{FF14,SKB15}. 
Also, \cite{KKBOP15} have studied the dynamo action in a broad range of Rossby 
numbers and found an anti-solar rotation pattern in slowly rotating stars. 
In our simulations, the slowest
rotating models with $\Omega=\Omega_{\odot}/2$, ~CZ03 and ~RC03, 
develop a large-scale magnetic field whose energy is several orders 
of magnitude smaller than the kinetic energy.  We did not obtain a regime 
at which the rotation pattern switches from solar to anti-solar. 

The models without a tachocline develop substantial latitudinal differential
rotation. Tilted, conical, iso-rotation contours are obtained for the models 
with relatively slower rotation (i.e., models ~CZ02 and ~CZ03).
The results reproduce remarkably well the differential rotation in the
solar convection zone, with the equator-to-$60^{\circ}$ latitude difference, 
$\chi_{\Omega} \sim 0.18$ (the same as in the Sun),
and also reproduce the near-surface shear layer (NSSL).

Because of the highly sub-adiabatic stable region, the simulations 
including the upper part of the radiative interior 
develop a tachocline, i.e., a rotational shear layer between the
radiative and convective zones. The model with faster rotation 
($\Omega=2\Omega_{\odot}$),  ~RC01, does not develop a NSSL. The
slower rotating models, which have a 
larger influence of the buoyancy force, RC02 ($\Omega=\Omega_{\odot}$)
and RC03 ($\Omega=\Omega_{\odot}/2$), do develop a NSSL.

The meridional circulation in the models with radiative zone show a 
poleward flow at latitudes $>30^o$ and a multicellular pattern radially 
distributed. This last characteristic resembles and theoretically supports 
recent helioseismology results. It is noteworthy, however, that to this 
date no global simulation has been able to accurately reproduce both, 
the solar differential rotation and the meridional circulation. The 
computational difficulties to simulate the processes at the upper part 
of the convection zone are perhaps the reason for the general lack 
of success.

Large-scale dynamo action is observed in all models. The
differential rotation, the dynamo 
growth rate and magnetic field topology depend on the Rossby number. 
The dynamo solutions include both steady and oscillatory dynamo regimes. 
The ratio between the magnetic and kinetic energies increases 
with the rotation rate, i.e., 
the smaller $\Ro$ the larger $e_{\rm M}/e_{\rm K}$ (see Table \ref{tbl.1}). 
Our results partially agree with those of \cite{Petit+08} since we 
observe a linear dependence of the mean magnetic energy with the rotation 
rate. Furthermore, in the fast rotating model with the tachocline, ~RC01, 
the magnetic energy is deposited in the poloidal 
magnetic field. Although most of this poloidal field is concentrated 
at the tachocline, this is reminiscent of the magnetic structure of 
fast rotating young stars,  whose observations indicate 
strong poloidal fields \citep{GDMHNHJ12}.

Using our simulations results, we have computed the inductive 
and dissipative terms of the mean-field induction equation.
We notice first that in both sets of dynamo models, 
the inductive term due to the radial shear is inversely 
proportional to the rotation rate.
The shear localizes mainly in the  boundary layers, i.e., 
the interface between the radiative and convective zones and in the
upper surface.
On the other hand, the inductive term due to the 
latitudinal shear in the simulations without the tachocline is 
proportional to $\Omega_0$. In the models RC01, RC02 and RC03, when 
the rotation is  fast,  $\partial_{\theta} \Omega$ peaks at the 
equator (RC01). When the rotation is slow, $\partial_{\theta} \Omega$  
peaks at the poles (RC03). 

The turbulent coefficients  
$\etat$ and $\alpha$ were computed using the FOSA approximation. 
We observe no dependence of the kinetic counterpart of $\alpha$, $\alpha_k$,  
on the Rossby number. However, the magnetic counterpart $\alpha_m$ (Eqs. 6 and 7) 
clearly anti-correlates with $\Ro$; i.e., the smaller $\Ro$, the larger the 
magnetic field and, thus, the larger the small-scale current helicity.
Consequently, the magnetic $\alpha$-effect,
computed here via the current helicity, could work as an inductive
term of the magnetic field generation. 
This partially agrees with the results of
\cite{VS14} in the sense that both, kinetic and magnetic parts of the
$\alpha$-effect may act as a source of magnetic field. 

In the oscillatory dynamo models without tachocline ~CZ01 and ~CZ02, 
the evolution 
of the field is consistent with dynamo waves propagating according
to the Parker-Yoshimura sign rule. It is not the case, however, for the
tachocline model, ~RC02, in which the migration of the field only in 
some regions can be explained by this rule.  Because the evolution 
of the field in this model is slow (on a time-scale of $\sim 30$ years), 
other factors like the meridional circulation and/or turbulent magnetic 
pumping, could be influencing the field transport. 

Though the FOSA 
is the simplest form of computing the mean-field coefficients,
in general terms our 3D simulation results give support to the 
mean-field dynamo theory as a formalism to explain solar and stellar
dynamos. A more sophisticated technique, like the test-field method 
\citep[e.g.,][]{SRSRC05,BRRS08}, could provide a better inference of 
these coefficients and thus allow a better analysis. 

The cycle period for the models without tachocline, ~CZ01 and ~CZ03,
is short (about $2$ years) compared to the solar cycle period. 
In the slow and fast rotating 
tachocline models, ~RC01 and ~RC03, the dynamo regimes are stationary.
However in the model ~RC02, with the solar rotation rate and the tachocline, 
the field is oscillatory with the period of about $30$ years. This time scale 
is comparable  with the solar cycle, however, unlike in the Sun, the 
toroidal magnetic field is symmetric across the equator.

In the RC models, strong toroidal magnetic field generated by dynamo is
concentrated
at or below the interface between the convective and the radiative layers. 
This indicates that the main source of the magnetic field is 
the radial shear at the tachocline. This effect has enormous influence on
the general evolution of the models. First, this strong magnetic field
modifies the mean-flow profiles via the Maxwell stresses; and second, because
the field is deposited in a convectively stable region, its evolution occurs 
on longer time-scales than in the models without the tachocline. 
Despite the field production is distributed  over the simulation
domain, the dynamo period is regulated by the diffusion time of 
the toroidal magnetic field at the deeper, more stable layers. 
The deep seated magnetic 
field seems to control the evolution of the magnetic field in the rest of the
convection zone. This result could explanain a long standing problem 
in mean-field modeling, i.e., the coexistence of a highly turbulent 
magnetic diffusivity,
of the order of $10^9$ m$^2$ s$^{-1}$, with the 22-year cycle period of 
the Sun. 

The value of the magnetic diffusivity, and so of the diffusion time,
in the stable layer is a product of the small scale motions generated by
MHD instabilities. We have studied the temporal 
evolution of the mean, axi-symmetric, 
as well as the non-axisymmetric kinetic and magnetic energy components. This temporal
evolution suggests that once the toroidal field is established due to the
strong radial shear, magneto-shear instabilities develop in the radiative
zone generating turbulent motions and magnetic fields.  The results
support the idea of a magnetic $\alpha$-effect as a source of poloidal
field. The turbulent motions, on the other hand, enhance the magnetic
diffusivity. Therefore, the system adjusts itself to the new quantities
until a new steady state is reached. The dynamics in the stable layer
depends on the Rossby number, slow rotating models exhibit more complex
behaviour and take longer to achieve steady state.  In model RC02 the
turbulent motions in the stable layer define the $\sim30$-year cycle
period. The solar model of tachocline of \cite{GCS10} uses a stratification
that results in slower convective, less penetrative,  motions. For that
reason the cycle period is longer that the one found here. The difference
in the cycle period between models with and without tachocline, as well
as the develpment of a magnetic $\alpha$-effect,  suggest
that the instabilities in the tachocline might be of significant importance
in the solar dynamo. 

One could ask why RC models do not exhibit the same short-period 
dynamo that is observed in the CZ models due to the latidudinal shear 
alone. The answer relies on the backreaction of the magnetic
field on the fluid dynamics via the Lorentz force. This influence can 
be easily noticed by comparing the profiles of $\Omega$ between the models
CZ02 and RC02. In numerical experiments, not presented here, we have 
verified that the strength of the toroidal magnetic field (and thus its 
influence on the flow) at the tachocline can be controlled by
varying different model ingredients. The radial shear source term
is particularly sensitive to the ambient state profile, $\Theta_e$. 
For instance, 
changes in parameters like the thickness of the transition between radiative and 
convective zones, the amplitude of the thermal oscillations or the
adiabaticity of radiative and convective layers, could result in different 
rotation and dynamo patterns.  The solar values of these 
parameters are not fully established for the Sun. 
We should also stress that a better understanding of 
the effects of the turbulent diffusion, in particular, 
the feedback of the magnetic field upon this diffusion is 
still necessary
\citep[e.g.,][]{dPLSGKL12,KRBKK14,dPLSGKL15}.

We conclude that tachoclines
play an important role in solar and stellar dynamos. Modelers should be 
careful when interpreting results obtained in simulations that
do not include this rotational shear layer. From this work, several new studies 
seem to be promising for the
understanding of the solar/stellar rotation and magnetism. For instance, 
a study of mean-flows and angular momentum balance for cool stars 
in the presence of magnetic fields could help to understand the cyclic
and non-cyclic stellar activity. Models without tachocline, which
evolve in short time-scales, and could be interpreted more easily, 
could help in the study of the origin of stochasticity 
and intermittency in turbulent flows. This will provide new insights 
on the physical origin of  Mounder-like solar minima. The results are also 
encouraging to the study of the rotation evolution and magnetism in young 
stars.  
\acknowledgments
This work was partly funded by FAPESP grants 2013/11679-4 (GG) and 
2013/10559-5 (EMGDP), CNPq grant 306598/2009-4 (EMGDP),  
NASA grants NNX09AJ85g and NNX14AB70G. GG also thanks
CNPq for travel support. PKS is supported by funding received from the 
European Research Council under the European Union's Seventh Framework 
Programme (FP7/2012/ERC Grant agreement no. 320375). The simulations
were performed in the NASA cluster Pleiades and the computing facilities 
of the Laboratory of Astroinformatics (IAG/USP, NAT/Unicsul) supported by
a FAPESP (grant 2009/54006-4). 

\bibliographystyle{apj}
\bibliography{bib}

%BEGIN
%END


%% in parentheses, after the keyword, but they will not be verified.

\clearpage

%% Here we use \plottwo to present two versions of the same figure,
%% one in black and white for print the other in RGB color
%% for online presentation. Note that the caption indicates
%% that a color version of the figure will be available online.
%%

%% This figure uses \includegraphics to scale and rotate the still frame

%% If you are not including electonic art with your submission, you may
%% mark up your captions using the \figcaption command. See the
%% No more than seven \figcaption commands are allowed per page,
%% so if you have more than seven captions, insert a \clearpage

%% Tables should be submitted one per page, so put a \clearpage before

%% Two options are available to the author for producing tables:  the
%% deluxetable environment provided by the AASTeX package or the LaTeX
%% table environment.  Use of deluxetable is preferred.
%%

%% Three table samples follow, two marked up in the deluxetable environment,
%% one marked up as a LaTeX table.

%% In this first example, note that the \tabletypesize{}
%% command has been used to reduce the font size of the table.
%% We also use the \rotate command to rotate the table to
%% landscape orientation since it is very wide even at the
%% Note also that the \label command needs to be placed

%% This table also includes a table comment indicating that the full
%% version will be available in machine-readable format in the electronic

\clearpage

%% If you use the table environment, please indicate horizontal rules using
%% Do not put multiple tabular environments within a single table.
%% The optional \label should appear inside the \caption command.

\clearpage

%% If the table is more than one page long, the width of the table can vary
%% from page to page when the default \tablewidth is used, as below.  The
%% individual table widths for each page will be written to the log file; a
%% maximum tablewidth for the table can be computed from these values.
%% The \tablewidth argument can then be reset and the file reprocessed, so
%% that the table is of uniform width throughout. Try getting the widths
%% from the log file and changing the \tablewidth parameter to see how
%% adjusting this value affects table formatting.

%% The \dataset{} macro has also been applied to a few of the objects to
%% show how many observations can be tagged in a table.

\end{document}